\documentclass[english,a4paper, aps, prb, citeautoscript, twocolumn, showpacs,superscriptaddress]{revtex4-1} 

\usepackage[english]{babel} 

\usepackage[T1]{fontenc}
\usepackage[utf8]{inputenc}
\usepackage[tbtags]{amsmath}
\usepackage{graphicx}
\usepackage[paperwidth=210mm,paperheight=297mm,centering,hmargin=1.8cm,vmargin=2.2cm]{geometry}

\usepackage{tikz}
\usetikzlibrary{external}
\usepackage{pgfplots}

\newcommand{\bra}{\langle}
\newcommand{\ket}{\rangle}
\newcommand{\Tr}{\textrm{Tr}}

\newcommand{\Las}{\text{L}}

\usepackage{siunitx}

\newcommand{\iim}{\mathrm{i}}

\newcommand{\G}{G^{(1)}}
\newcommand{\Gtil}{\tilde{G}^{(1)}}
\newcommand{\gtil}{\tilde{g}^{(1)}}
\newcommand{\gttil}{\tilde{g}^{(2)}}
\newcommand{\Gt}{G^{(2)}}

\newcommand{\GtX}{G^{(2\text{X})}}
\newcommand{\g}{g^{(1)}}
\newcommand{\gt}{g^{(2)}}
\newcommand{\gtX}{g^{(2\text{X})}}

\newcommand{\Ep}{E^{(+)}}
\newcommand{\vEp}{\vec E^{(+)}}
\newcommand{\upDelta}{\Delta}

\newcommand{\ta}{t_\mathrm{A}}\newcommand{\tb}{t_\mathrm{B}}%
\newcommand{\ra}{r_\mathrm{A}}\newcommand{\rb}{r_\mathrm{B}}%

\newcommand{\dd}{\mathrm{d}}

\begin{document}

\title{Correlation functions with single photon emitters \\
under noisy resonant continuous excitation}
\author{E. Baudin}
\email{emmanuel.baudin@lpa.ens.fr}
\affiliation{Laboratoire Pierre Aigrain, Ecole normale sup\'erieure, PSL University, Sorbonne Universit\'e, Universit\'e Paris Diderot, Sorbonne Paris Cit\'e,  CNRS, 24 rue Lhomond, 75005 Paris France}
\author{R. Proux}
\affiliation{Laboratoire Pierre Aigrain, Ecole normale sup\'erieure, PSL University, Sorbonne Universit\'e, Universit\'e Paris Diderot, Sorbonne Paris Cit\'e,  CNRS, 24 rue Lhomond, 75005 Paris France}
\author{M. Maragkou}
\affiliation{Laboratoire Pierre Aigrain, Ecole normale sup\'erieure, PSL University, Sorbonne Universit\'e, Universit\'e Paris Diderot, Sorbonne Paris Cit\'e,  CNRS, 24 rue Lhomond, 75005 Paris France}
\author{Ph. Roussignol}
\affiliation{Laboratoire Pierre Aigrain, Ecole normale sup\'erieure, PSL University, Sorbonne Universit\'e, Universit\'e Paris Diderot, Sorbonne Paris Cit\'e,  CNRS, 24 rue Lhomond, 75005 Paris France}
\author{C. Diederichs}
\affiliation{Laboratoire Pierre Aigrain, Ecole normale sup\'erieure, PSL University, Sorbonne Universit\'e, Universit\'e Paris Diderot, Sorbonne Paris Cit\'e,  CNRS, 24 rue Lhomond, 75005 Paris France}

\date{\today}
\begin{abstract}
To characterize the statistics and indistinguishability of a source, it is common to measure the correlation functions of the emitted field using various interferometers. 
Here, we present a theoretical framework for the computation of the correlation functions of a two-level system that is resonantly driven by a realistic noisy cw excitation laser. 
Analytic expressions of the first and second-order  auto-correlation functions are obtained where the various contributions of the noisy excitation source are correctly taken into account. 
We predict that, even in the low power regime, the noise source has a strong influence on the two-level system dynamics, which is not anticipated by simpler models. 
The characterization of photon indistinguishability in the pulsed excitation regime is usually done by measuring the value of the zero-delay intensity correlation obtained with a Hong-Ou-Mandel interferometer. We show that this figure is irrelevant in the cw excitation regime and we introduce the coalescence time window, a figure of merit based on a probabilistic interpretation of the notion of photon indistinguishability. We finally use the coalescence time window to quantify how noisy cw excitation influences photon indistinguishability. 
\end{abstract}
\maketitle
\section{Introduction}
\label{s:intro}

The resonance fluorescence, i.e. the emission of photons by a two-level system irradiated by a resonant laser field, has attracted much attention since an electronic transition between two well-defined energy levels results in the emission of single photons. The investigation of such emission dynamics under resonant pumping started experimentally in single atoms \cite{Kimble1977} or ions \cite{Diedrich1987} and was then expanded in other systems such as single molecules \cite{Basche1992}, color centers in diamond \cite{Bouri2000} or semiconductor quantum dots \cite{Muller2007}. In these latter systems, a strong interest has been devoted to the improvement of single photon emission in terms of indistinguishability properties and collection efficiencies\cite{He2013,Somaschi2016,Senellart2017} for the development of integrated indistinguishable single photon sources for quantum information applications.

The statistics and indistinguishability properties of the emitted photons are experimentally investigated by measuring the second-order intensity auto- and cross-correlation functions, $\gt$ and $\gtX$, in a Hanbury-Brown and Twiss experiment \cite{HBT} and a Hong-Ou-Mandel interferometer \cite{HOM}, respectively. These properties are closely linked to the intrinsic characteristics of the emitter: the lifetime $T_1$ which is accessible by time-resolved photoluminescence experiments, and the dephasing time $T_2$ which is evaluated by measuring the first-order field auto-correlation function $\g$ in a Fourier transform spectroscopy experiment (with a Michelson interferometer for example).
All the involved correlation functions are routinely used and well established in the case of a two-level system excited by a non-resonant laser, and their analytical expressions depend on the pumping rate, $T_1$ and $T_2$ \cite{Loudon}. As far as photon indistinguishability is concerned, the zero-delay value of the $\gtX$ correlation function gives the degree of indistinguishability which is intrinsically linked to the ratio $T_2/2T_1$ under pulsed excitation \cite{Bylander2002,Santori2002}. However, in the case of continuous excitation, the time constants of the emitter govern the width of the correlation function whereas its zero-delay value is mainly imposed by the time response of the detection system \cite{Legero2004,Halder2008}. This makes the usual characterization of the photon indistinguishability inappropriate under continuous wave (cw) excitation.

In this context, we presented an experimental study on photon indistinguishability where a more appropriate figure of merit, the coalescence time window (CTW), was introduced to measure the indistinguishability of a cw single photon source \cite{Proux2015}. In the resonant Rayleigh scattering (RRS) regime (also referred as the Heitler regime) where the spectrum is dominated by an elastic component characterized by the wavelength and the linewidth of the laser \cite{Nguyen2011}, we have shown that the photon indistinguishability is governed by the coherence time of the laser, ensuring the generation of highly indistinguishable single photons in terms of CTW\cite{Proux2015}. In this experimental study, the correlation functions evaluated for a resonantly-driven two-level system \cite{Scully1997} were used to analyze the results but a precise theoretical study of the contribution of the excitation source is still necessary since the computation of the correlation functions of the emitted field inevitably depends on the characteristics of the excitation source in the RRS regime.

With nowadays laser sources, the laser coherence times lie in the few tens of microseconds range, meaning that the minimum HOM interferometer arm length difference should be at least of several kilometers in order to avoid the beating of one-photon interferences in the RRS regime. Moreover, the basic hypothesis at stake for HOM result analysis is that the emission of photons at sufficiently long delays are totally uncorrelated with each others so that the source could be replaced by two independent identical emitters. This hypothesis is clearly not fulfilled under cw resonant excitation due to the memory effect introduced by the phase of the exciting laser field. Therefore, to restore these fundamental requirements, we used a noisy laser source which clears this memory effect and ensures the validity of the HOM two-photon interference experiment. However, in doing so, the dynamics of the two-level system is itself impacted, which leads to other modifications on the interferogram: 
shorter effective relaxation constants and effective blurring of the interferograms due to time-averaging (in the case of an electronically driven noisy source which is under the scope of this paper). 
In this paper, we define the various regimes of the two-level system dynamics under noisy continuous resonant excitation (sect.~\ref{s:fluctsource}) and address the question of the calculation of multiple-time correlation functions of the field emitted by a two-level system (sect.~\ref{s:Firstandsecond}). Finally, we investigate the HOM interferometer response (sect.~\ref{s:propagation}). In this context we discuss in details the introduction of the CTW as a figure of merit for the photon indistinguishability in the cw regime, and we discuss the effect of a noisy driving on the CTW and zero-delay intensity cross-correlation measurements.


\section{Effects of a noisy driving field on the dynamics of a two-level system}
\label{s:fluctsource}

\subsection{The two-level system dynamical equations} 

\subsubsection{The Bloch equations in the fixed frame - Liouville equation}

Let us consider a two-level system governed by a Hamiltonian $H_0$. 
The dynamics of the density matrix $\hat \rho$ describing the state of this system in the laboratory frame is given by the Liouville equation:
\begin{equation}
i \hbar \partial_t \hat \rho = [ H_0, \hat \rho],
\end{equation}
where $H_0=\hbar \omega_0 {S_+} {S_-}$, $\hbar \omega_0$ is the transition energy of the two-level system, and ${S_+}$ (${S_-}$) is the highering (lowering) ladder operator of the two-level system. 
An additionnal time-dependent Hamiltonian $H_1 (t)$ allows accounting for the resonant driving of the two-level system. 
If this term is purely resonant, it takes the form $H_1 (t) = \hbar \Omega_1 {S_+} \cos(\omega_0 t) + \mathrm{h.c.}$, where $\Omega_1$ is the (Rabi) angular frequency associated to the driving amplitude. 
If this term is only partially resonant it takes the form $\hbar (\bar{\Omega}_1+ \delta \Omega_1 (t)) {S_+} \cos((\omega_0 +\Delta \omega) t + \int_0^t \delta \omega (t') \dd t') + \mathrm{h.c.}$, where $\delta \Omega_1 (t)$ and $\delta \omega (t)$ are the time-flucutating coupling amplitude and angular frequency of the driving field, and $\bar{\Omega}_1$ and $\Delta \omega$ are the secular coupling amplitude and driving angular frequency, respectively.

\subsubsection{The fluctuating excitation field}
\label{s:deffluct}

We consider the case where fluctuations $\delta \Omega_1 (t)$ and $\delta \omega (t)$ result from the fluctuations of the excitation field which we define as

\begin{equation}
E(t)=\left( \bra E \ket +\delta E(t) \right)
\cos \left( \int_0^t \omega_0 + \delta \omega (t')\  dt' \right),
\end{equation}
where $\delta E(t)$ and $\delta\omega(t)$ are the time-dependent fluctuating amplitude and angular frequency of the field, respectively.

In this work, the fluctuations are characterized by their first-order correlation functions:
\begin{equation}
 \begin{cases}
	\overline{\delta E (t) \delta E (t+\tau)}, \\
	\overline{\delta \omega (t) \delta \omega (t+\tau)  }, \\
	\overline{\delta E (t) \delta  \omega (t+\tau)  },
     \end{cases}
\label{e:CorrFunc}
\end{equation}
where ovelined quantities are averages over the noise realization, which corresponds to averaging over time $t$. Those are assumed to be monoexponential laws fully characterized by their root-mean-squared (rms) amplitude at zero delay and their correlation time $\tau_C$.
The respective probability density functions are assumed to be Gaussians, e.g., $p(\delta \omega) \propto \exp (-\delta \omega^2/2\overline{\delta \omega^2})$. 
These fluctuations can be inherent to the excitation or created artificially by driving a laser diode with a laser coherence controller (LCC) as in ref. \cite{Proux2015}.

\subsubsection{Bloch equations in the rotating frame at the instantaneous angular frequency}
\label{s:rotframe}
In the absence of fluctuations in the driving field, the usual procedure consists in using the rotating wave approximation (RWA) in the frame rotating at the laser frequency. This leads to the well-known Bloch-Liouville equation: 
\begin{equation}
i\hbar \partial_t \tilde{\rho} =
[\tilde H , \tilde{\rho}]  + \mathbf{R} \mathbf{[} \tilde{\rho} \mathbf{]},
\end{equation}

where $\tilde H = \hbar (\Delta \omega \tilde {S_+} \tilde {S_-} + \frac{\Omega_1}{2} (\tilde {S_+} + \tilde {S_-}))$ and 
the term $\mathbf{R} \mathbf{[} \tilde{\rho} \mathbf{]}$ is a Markovian relaxation term describing the effect of the environnement on the two-level system dynamics. In the usual weak driving case ($|\Omega_1|\ll|\Omega_0|$) these terms are the longitudinal ($T_1$) and transverse ($T_2$) relaxations.
In this frame, after the RWA, the time-dependence of the driving disappears and these equations can be solved analytically. 

In the case of a noisy driving, we can still perform the RWA in the frame rotating at the instantaneous laser angular frequency $\omega_0 +\Delta \omega+\delta \omega (t)$, but the resulting modified Bloch-Liouville equation is now time-dependent:
\begin{equation}
i\hbar \partial_t \tilde{\rho} =
[\tilde H , \tilde{\rho}] + [\delta \tilde H(t) , \tilde{\rho}]  + \mathbf{R} \mathbf{[} \tilde{\rho} \mathbf{]}
\end{equation}

 with $\delta \tilde H(t) = \hbar \delta \omega (t) (\tilde {S_+} \tilde {S_-}) + \hbar \frac{\delta \Omega_1 (t)}{2} (\tilde {S_+} + \tilde {S_-})$, and the relaxation operator $\mathbf{R} \mathbf{[} \tilde{\rho} \mathbf{]}$ gets an extra contribution induced by fluctuations (see appendix \ref{a:BPP}). Approximations are required to solve this equation in specific cases as we shall see in the next section. 

\subsubsection{Correlation functions}

The single photon characteristics of a two-level emitter is characterized by its first and second order correlation functions which involve two- and four-time temporal correlators of the form $\bra {S_+} (t_2)\, {S_-}(t_1)\ket$ or $\bra {S_+} (t_4)\, {S_+} (t_3)\, {S_-}(t_2)\, {S_-}(t_1)\ket$, respectively. In general, these times can all be different. 
For a time-independent Markovian dynamics, the quantum regression theorem is the tool of choice to compute the correlation functions\cite{Lax1963, Lax1967}. When multiple-time correlation functions need to be evaluated, it can be more convenient to use the superoperator formalism to apply the quantum regression theorem. A straightforward method which is out of the scope of this work.  
In general, the scattered light correlation functions involve the driving source correlation function. 
Therefore, the computation of the correlation functions requires a careful handling of the two-level system dynamics under noisy resonant conditions. 


\subsection{The various regimes of driving}

Several regimes can be distinguished, depending on the relative importance of the correlation time $\tau_C$, the rms amplitudes $\sqrt{\overline{\delta \omega^2}}$, $\sqrt{\overline{\delta \Omega_1^2}}$ of the fluctuation driving terms, the average driving term amplitudes $\bar \Omega = \sqrt{\Delta \omega^2+ \bar \Omega_1^2}$ and the relaxation times $T_1$ and $T_2$.

\subsubsection{Monte Carlo or not?}


In general, the fluctuating Bloch-Liouville equation can be solved if the relaxation rates $T_1$ and $T_2$ are short compared to the coherence time $\tau_C$ (pseudo-adiabatic regime) or if the phase accumulations  due to fluctuations $\sqrt{\overline{\delta \omega^2}} \tau_C$ and $\frac {\overline \Omega }{\overline E }  \sqrt{\overline{ \delta E^2 }} \tau_C $ are smaller than 1 (Bloch-Purcell-Pound regime, BPP). 
If one of these conditions is not met, the dynamics can only be solved by Monte-Carlo simulations and post-averaging over the noise realization. 

If these conditions are met, simplifications exist to solve the fluctuating Bloch-Liouville equation analytically in the instantaneous rotating frame. 
Before looking in detail at the BPP and pseudo-adiabatic cases, let us remark that the correlation functions of interest (which are measured experimentally) are expressed in the laboratory frame. 
It is then necessary to handle carefully the rotating frame transformation.

\subsubsection{Rotating frame blurring}
\label{ss:rfblurring}

The rotating frame transformation naturally conveys a fluctuating dephasing term between the laboratory frame and the frame rotating at the instantaneous laser frequency. The corresponding accumulated phase describes a Brownian motion on a circle which leads, after averaging over noise realizations, to an effective mono-exponential relaxation with a decay rate $\Gamma_L = \overline{\delta \omega^2} \tau_C$ (see appendix \ref{a:fluctframe}). We call the corresponding dephasing-induced relaxation "rotating frame blurring" and the associated timescale $T_L=\Gamma_L^{-1}$ is the laser coherence time in reference to the driving laser used in quantum optics. Note that this coherence time is specific to the driving source and can be measured using a Michelson interferometer. 

The rotating frame blurring is uncorrelated with the two-level system evolution in the laser frame as long as the laser coherence time $T_L$ and the auto-correlation timescale $\tau_C$ are sufficiently different. This condition is met if the characteristic phase accumulation due to fluctuations $ \sqrt{\overline{\delta \omega^2}} \tau_C$ is smaller than 1. 

The passage from the laboratory frame to the rotating frame at the instantaneous angular frequency of the laser is defined in the density matrix formalism as $\tilde{\rho}(t)=R_{t_{0}\rightarrow t} \rho(t) R^\dagger_{t_{0}\rightarrow t}$, where $R_{t_{0}\rightarrow t}$ is the rotation operator at the instantaneous laser angular frequency and $t_{0}$
is the synchronization instant between the two reference frames. Consequently, $R_{t_{0}\rightarrow t_{0}}=1$.
From this definition, we define the expression of operators in the rotating
frame as $\tilde{O}(t)=R_{t_{0}\rightarrow t} O (t) R_{t_{0}\rightarrow t}^{\dagger}$.

When this rotating frame transformation is applied to highering and lowering dipolar operators ${S_{\pm}} (t)$, a remarkably simple result is obtained:
\begin{equation}
{\tilde{S}_{\pm}}(t_{1})={S_{\pm}}(t_{1})e^{\mp i\phi_{t_{0}\rightarrow t_{1}}},
\label{e:SinRotFrame}
\end{equation}
\noindent where $\phi_{t_{0}\rightarrow t_{1}} = \int_{t_0}^{t_1} \delta \omega (t) dt$ is the accumulated phase between $t_{0}$ and $t_{1}$.

Applying this result to the  multiple-time correlator expression in the laboratory frame 
leads to
\vspace{-0.1cm}
\begin{equation}
\left\langle {S_+}(t_{2}){S_-}(t_{1})\right\rangle =
e^{i(\phi_{t_{0}\rightarrow t_{1}}-\phi_{t_{0}\rightarrow t_{2}})}
\big\langle {\tilde{S}_{+}}(t_{2}){\tilde{S}_{-}}(t_{1})\big\rangle,
\end{equation}

which reduces to
\begin{equation}
\left\langle {S_+}(t_{2}){S_-}(t_{1})\right\rangle =
e^{-i \phi_{t_{1}\rightarrow t_{2}}}
\big\langle {\tilde{S}_{+}}(t_{2}){\tilde{S}_{-}}(t_{1})\big\rangle.
\end{equation}

Generalization to $n>2$-time correlators is straightforward if phase accumulating periods do not overlap. 

\subsubsection{Relaxation in the rotating frame}


The influence of a fluctuation term with a vanishingly short correlation time in the Bloch equation is well-known since the pioneering work by Bloembergen, Purcell, and Pound (BPP) on liquid state NMR\cite{Bloembergen1948, Abragam1962} in 1948. 
If the average phase accumulation $\overline {\Omega} \tau_C$ during the correlation time is much smaller than 1, 
it results in an effective static relaxation term following the example in NMR of the derivation of longitudinal and transverse relaxation of a single spin due to a time-varying magnetic field produced by its moving first neighbors in the ``non-viscous liquid'' limiting case (see appendix~\ref{a:BPP}). Similarly, if this condition is not met, but if  $\sqrt{\overline{\delta \omega^2}} \tau_C \ll 1$ and $\frac {\overline \Omega }{\overline E }  \sqrt{\overline{ \delta E^2 }} \tau_C \ll 1$, then the more general ``viscuous liquid'' regime is reached and similar relaxation operators can be found analytically\cite{Tomita1958}. 
Note that in both cases, BPP formulas can only be used in the rotating frame at the instantaneous laser angular frequency $ \omega_0 + \delta \omega (t)$ in which the Bloch equations are time-independent and can be explicitely solved. To obtain results in the laboratory frame requires to take into account the rotating frame blurring effect previously described. 


\subsubsection{Pseudo-adiabatic evolution}

Another limiting case which can be handled analytically is the quasi-adiabatic evolution which occurs if relaxation is much faster than the correlation time, \textit{i.e.}, $T_2, T_1 \ll \tau_C$. In this case, the dynamics can be handled as if no fluctuating term was present except that averaging over the statistics of fluctuations has to be done afterwards prior to rotating frame blurring. 

In the pseudo-adiabatic regime, the fluctuating term $ \delta H(t) $ variation is slow compared to the typical two-level system relaxation timescale $T_2$ and Bloch equations can be considered as coupled quasi-static ordinary differential equations. Averaging over the different realizations of the fluctuations has to be taken into account once the fluctuationless correlation function in the laboratory frame is obtained, and, at the level of the optical Bloch equations, we are left with the usual static case.

\subsubsection{Experimental regime}

\begin{figure}[!ht]
\begin{center}
\includegraphics[width=8cm]{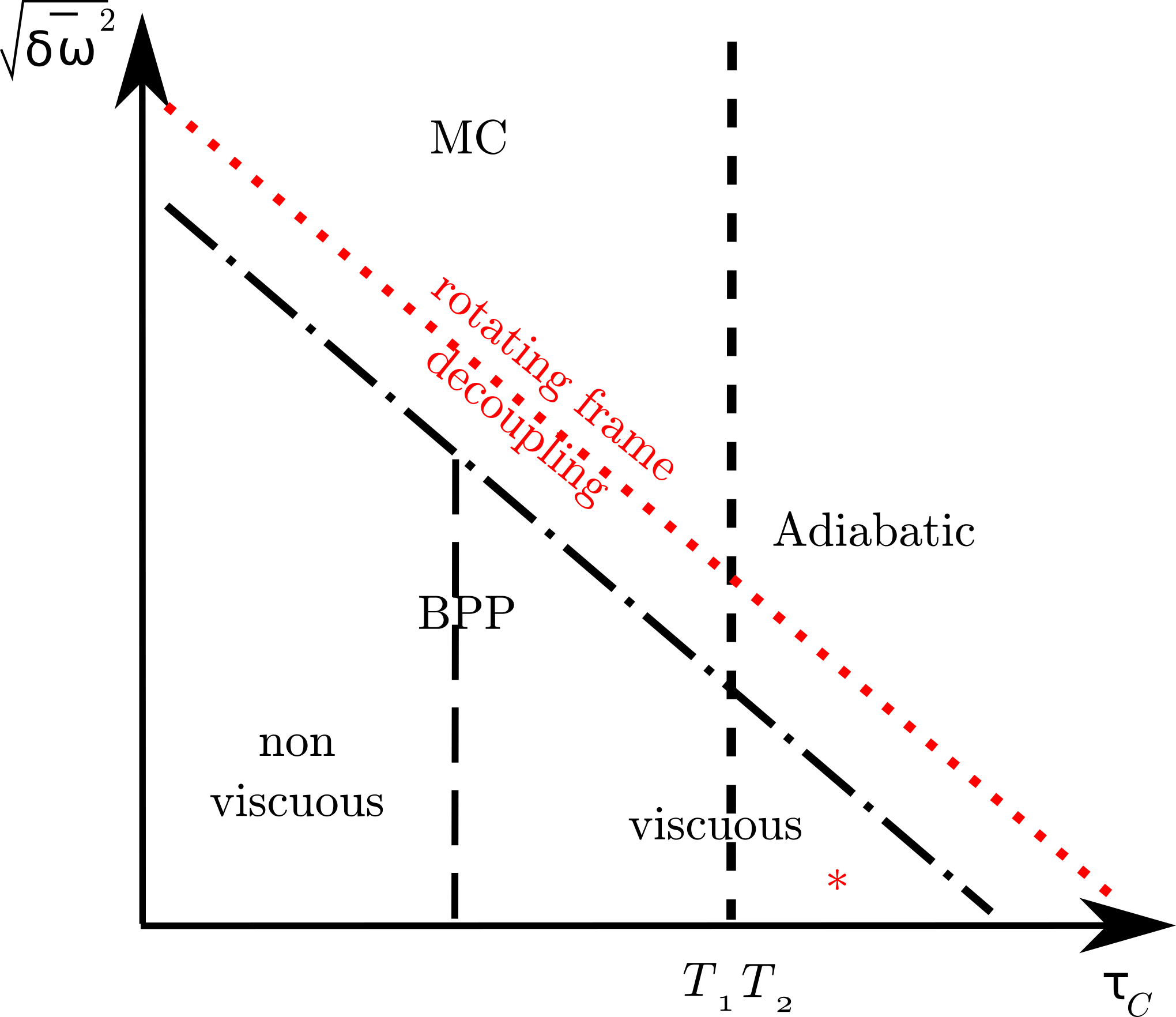}
\caption{\label{Fig2} Diagram of the different evolution regimes depending on the angular frequency fluctuation characteristics of the continuous driving: (i) x-axis: coherence time $\tau_C$ of the fluctuation (ii) y-axis: square-root of the fluctuation $\overline{\delta \omega ^2}$. Both axes are implicitly in logarithmic scales. The diagonal dotted-dashed line delimitate the left bottom side where BPP approximation can be used either in the viscous or non-viscous regime. The vertical dashed line delimitate the region where the pseudo-adiabatic approximation is valid. In the rest of the diagram, the correlation statistics can be reached through Monte Carlo (MC) simulations. The red dotted line represents the region where relaxations processes decoupling is valid (see section~\ref{s:rotframe}). This line collapses with the BPP frontier if the field fluctuation amplitudes are smaller. Note that in the region common to BPP and adiabatic regimes, BPP relaxation can be neglected. The red star spots refer to the experimental conditions investigated in reference~\cite{Proux2015}.
}
\end{center}
\end{figure}

Figure \ref{Fig2} summarizes the different regimes depending on the amplitude and correlation time of the fluctuations of the source. The diagram presents three regions of interest: The lower triangle at short correlation time and amplitude fluctuations is the BPP regime where the fluctuations result in effective extra relaxation terms in the Bloch equations and a rotating frame blurring effect. 
The right rectangle at long correlation times is the adiabatic regime region, where post-averaging over realizations is responsible for the shape of correlation response functions. Note that, in the adiabatic regime, rotating frame decoupling dashed line delimit the region (at small fluctuation amplitude) where adiabatic averaging can be done independently of post-averaging over realizations. Finally, the upper-left triangle at short correlation time but large fluctuation amplitude cannot be computed analytically and require Monte Carlo simulation.

In reference~\cite{Proux2015} in which the emitter is a self-assembled InGaAs quantum dot, $T_1$ and $T_2$ are subnanosecond and the laser fluctuations are driven by an external LCC that has a bandwidth limited to 250 MHz by the electronics. 
Using a Hanbury Brown and Twiss interferometer\cite{Proux2015}, we obtain the characteristic correlation time of the laser field  $\tau_C \sim \si{4}{ns}$ and the driving field fluctuation amplitude $ \overline{ \delta E^2 } /\overline{ E^2 } \simeq 3 \% $.
Consequently $\tau_C > T_1,\, T_2$, and the amplitude is chosen such that decoupling occurs on the HOM interferometer path difference, i.e. $1/\Gamma_L <\Delta t = 43.5 ns $ (9 m propagation in an optical fiber), whereas the rotating frame decoupling condition is given by $\Gamma_L \tau_C \sim 0.1 \ll 1$. 
With this fluctuating source, the regime of driving is pseudo-adiabatic. In this regime, when the rotating frame decoupling condition is met, the optical Bloch equation is unperturbed in the instantaneous rotating frame, and, as we shall see in the next section, in the weak driving limit the interferogram can be understood using a straightforward unperturbed cw-driving interpretation. Consequently, the pseudo-adiabatic regime is of particular relevance for the characterization of two-level system emitters.

We note that this framework can also be used to understand the symmetric situation where a noise-free excitation is used, but the two-level system is subject to energy fluctuations due to the evolution of its surrounding. Note that in this configuration, there is no rotating frame averaging effect. If energy fluctuations are small enough, their effect can be classified depending on their frequency content in non-viscuous (BPP) relaxation (high frequencies), viscuous relaxation (medium frequencies) and pseudo-adiabatic averaging (low frequencies). 

Having now classified the various dynamical regimes of a two-level system under resonant noisy excitation, let us now turn to the determination of the associated two-time correlation functions.

\section{First and second order correlation functions}
\label{s:Firstandsecond}

First and second order normalized correlation functions $g^{(1)}$ and $g^{(2)}$ are the tools of choice to characterize field coherence and single-photon caracteristics of a radiated e.m. field. 
They are measured using a Michelson inteferometer and a Hanbury-Brown and Twiss interferometer, respectively. If the two-level system is either non-resonantly excited or resonantly excited with a monochromatic excitation, the theoretical calculation of these correlation function is seamless. 
It is more complex however under noisy resonant driving due to rotating frame blurring and pseudo-adiabatic averaging over noise realizations.


\subsection{Rotating frame blurring effect on $g^{(1)}$ and $g^{(2)}$}

We define the normalized correlation functions $g^{(1)} (\tau)=G^{(1)} (\tau)/G^{(1)} (0)$ and $g^{(2)} (\tau)=G^{(2)} (\tau)/G^{(2)} (\infty)$ from the non-normalized correlation functions $G^{(o)} (\tau)$, where $o$ is the correlation function order. Let us first consider the non-normalized first-order correlation function $\G$:
\begin{equation}\big\langle {S_+}(t+\tau){S_-}(t)\big\rangle =e^{i\phi_{t\rightarrow t+\tau}}\big\langle {\tilde{S}_{+}}(t+\tau){\tilde{S}_{-}}(t)\big\rangle \end{equation}
Hence, after averaging over the fluctuations in the Brownian motion limiting case and normalization:
\begin{equation}
g^{(1)}(\tau)=e^{-\Gamma_{\Las}|\tau|}e^{i\overline{\omega}_{\Las}\tau} \ \tilde{g}^{(1)}(\tau)\label{g1}.
\end{equation}

For $\Gt$, phase accumulation cancels and, assuming $\tau>0$:
\vspace{-0.2cm}
\begin{multline}
 \left\langle {S_+}(t){S_+}(t+\tau){S_-}(t+\tau){S_-}(t)\right\rangle  =  \\
\big\langle {\tilde{S}_{+}}(t){\tilde{S}_{+}}(t+\tau){\tilde{S}_{-}}(t+\tau){\tilde{S}_{-}}(t)\big\rangle
\end{multline}

so that
\vspace{-0.1cm}
\begin{equation}\gt(\tau)=\gttil(\tau).
\end{equation}

\subsection{Adiabatic averaging effects}

In the adiabatic case, averaging over excitation field fluctuations prior to rotating frame blurring has to be taken into account. The general expression reads

\begin{multline}
\overline{\tilde G^{(1,2)}}(\tau) = \iiiint \dd E_1 \ \dd E_2 \ \dd \delta \omega_1 \  \dd \delta \omega_2 \\ 
 p(E_2,\delta \omega_2, t_2;\  E_1,\delta \omega_1, t_1) 
\tilde G^{(1,2)}_{E_2,\delta \omega_2,\tau |E_1, \delta \omega_1,0},
\label{eq:pseudoadia}
\end{multline}

where $\tilde G^{(1,2)}_{E_2,\delta \omega_2,\tau |E_1, \delta \omega_1,0}$ is computed using (for example) the quantum regression theorem with driving parameters $(E_2, \delta \omega_2)$ while the equilibrium density matrix is computed using parameters $(E_1, \delta \omega_1)$. \\

The two-level system dynamics is characterized by a saturation parameter $s=\frac{\Omega_1^2 T_1 T_2}{1+ (\Delta \omega T_2)^2}$, where $\hbar \Omega_1 = d E$ and $d$ is the dipole amplitude of the considered transition. If $s < 1$, the two-level system is characterized by a linear response to driving and the system is in the weak driving regime. Conversely, if $s \geq 1$, the 2-level system is in the strong driving regime and is characterized by various non-linear response signatures such as Mollow triplet or Rabi oscillations. 
The complex pseudo-adiabatic expression (\ref{eq:pseudoadia}) can be simplified in several limiting cases, in particular in the weak driving regime $s < 1$.

Before considering those cases, let us note that in the pseudo-adiabatic regime
$$
\sqrt{\overline{ \delta \omega^2 }} = \sqrt{\frac{\Gamma_L}{\tau_C}} \ll \frac{1}{T_1},\frac{1}{T_2},
$$
i.e., typical energy fluctuations of the laser are much smaller than the radiative linewidth so that energy fluctuations can be neglected.

\subsubsection{Long correlation driving}

If the driving correlation time $\tau_C$ is much longer than $T_{1,2}$ which corresponds to the characteristic timescales of the $\tilde G^{(1,2)}$ decay, we can consider that the field characteristics are frozen on each realization of the experiment so that 
$p(E_2,\delta \omega_2, t_2;\  E_1,\delta \omega_1, t_1) \simeq p(E_1,\delta \omega_1) \delta (E_2-E_1) \delta (\delta \omega_2-\delta \omega_1)$. Consequently, 

\begin{equation}
\overline{\tilde G^{(1,2)}}(\tau) = \iint \dd E \  \dd \delta \omega \ 
p(E,\delta \omega) \ \tilde G^{(1,2)}_{E,\delta \omega}.
\end{equation}

For weak driving, linear response theory allows to further simplify this expression since
$\tilde G^{(1,2)}_{E,\delta \omega}$ is well reproduced by its Taylor expansion up to second order. 
After integrating over $E$ and $\delta \omega$, the following expression is obtained

$$\overline {\tilde G^{(1,2)}_{E, \delta \omega } } \simeq
\frac{1+C}{2} \tilde G^{(1,2)}_{\sqrt{\overline{ E^2 }},\sqrt{ \overline{ \delta \omega^2 }}}
+
\frac{1-C}{2} \tilde G^{(1,2)}_{\sqrt{\overline{ E^2 }},-\sqrt{\overline{ \delta \omega^2 }}}$$
where $C=\frac{\overline{ E \delta \omega }}{\sqrt{\overline{ E^2 }}\sqrt{\overline{ \delta \omega^2 }}}$ due to the symmetric response of $\overline {\tilde G^{(1,2)}}$ to $E$ and $\delta \omega$.


Neglecting the driving energy fluctuation, we get the simple results

$$\overline {\tilde G^{(1,2)}_{E, \delta \omega } } \simeq \tilde G^{(1,2)}_{\sqrt{\overline{ E^2 }},0}.$$

\subsubsection{Short correlation driving}
\label{ss:shortcorr}

In practical uses of a noisy source, the interferogram can be recorded on a timescale which is comparable to $\tau_C$ such as in ref \cite{Proux2015}. In the weak driving case, by using Taylor expansion, it is possible to obtain a long but explicit expression in terms of the correlation functions of the driving field multiplied by correlation functions at remarkable fields. 

In the case of the first-order auto-correlation function, the calculation is dramatically simplified in the weak driving regime. 
In this case $\Gtil_{E_2,\tau |E_1,0} = \alpha^{-2} E_2 E_1 \gtil (\tau)$, where $\alpha$ is a constant (defined in section \ref{ssec:Expr_HOM}), and averaging over driving field energy and amplitude fluctuations leads to 

$$
\overline{\Gtil_{E_2,\tau |E_1,0}} = \alpha^{-2}
\left( \overline {E}^2 + \overline{\delta E (t) \delta E (t+\tau)} \right) \gtil (\tau).
$$

The normalized $\overline {\gtil}$ is obtained by dividing the latter expression by $\overline {\Gtil_{E, \delta \omega } } (0)$, and we obtain

\begin{equation}
\overline {\gtil} (\tau) = \gtil (\tau) \left( 1-\frac{Q^{-2} (1-e^{-\tau/\tau_C})}{1+Q^{-2}} \right),
\label{eq-simple}
\end{equation}
where $Q^{-2}=\overline{\delta E^2}/\overline{E}^2$. $Q$ can be seen as the quality of the cw source: the noisier the source, the smaller is $Q$.

\begin{figure}[h!]
\centering
\includegraphics[width=8cm]{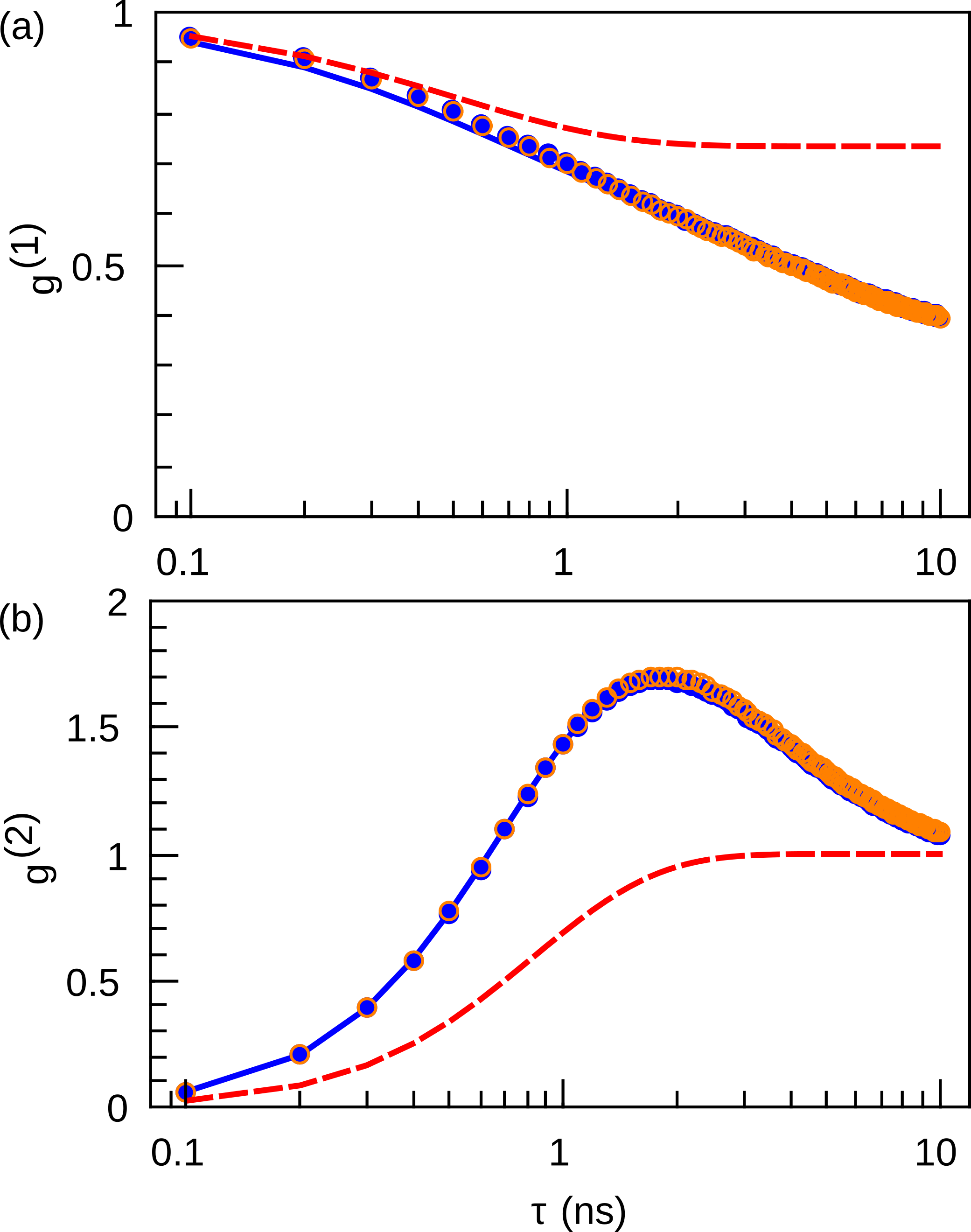}
\caption[graph]{ $\overline \gtil$ (a) and $\overline \gt$ (b) in the weak driving limit obtained by numerical integration of the Liouville equations taking into account pseudo-adiabatic averaging. The simulations are represented by symbols and are computed for a resonant excitation and for $\tau_C= 4 \si{ns}$, $T_1= 0.34 \si{ns}$ , $T_2= 0.5 \si{ns}$, $\overline \Omega=0.1 \si{rad/ns}$, and consequently a saturation parameter $s=1.7.10^{-3}$ (weak driving).
In both cases, $\sqrt{\overline{\delta \Omega^2}} = \sqrt{\overline{\delta \omega^2}} = 0.1 \si{rad/ns}$ which corresponds to $Q=1$.
The rotating frame blurring is not included in the calculation.
Both $\overline \gtil$ and $\overline \gt$ simulations have been done with a correlation factor between driving field amplitude and energy fluctuations $\epsilon=0$ (circles) and $0.8$ (dots).  
The fluctuationless theories are represented as dashed lines and fail at describing the simulations. The weak driving theory is represented as thick lines and reproduces accurately the simulations. 
}
\label{fig:Weak_driving}
\end{figure}

Figure~\ref{fig:Weak_driving} represents $\overline {\gtil} (\tau)$ (panel a) and $\overline \gt (\tau)$ (panel b) obtained by numerical integration in the weak driving limit. 
In both cases, we observe that driving field fluctuations indeed modify qualitatively the correlation response function. Furthermore, in the presence of driving field fluctuations, we observe that correlation response functions are independent of cross-correlation $\epsilon$ between energy and amplitude fluctuations of the noisy driving. We can conclude that, as pseudo-adiabatic relaxation results from the field amplitude fluctuations alone whereas rotating frame blurring results from energy fluctuations alone, in the weak coupling regime, the two mechanisms are independent even in the presence of cross-correlation between the two driving field fluctuations.
Equation (\ref{eq-simple}) captures quantitatively the first order correlation function $\overline {\gtil} (\tau)$ in the weak coupling regime (fig.~\ref{fig:Weak_driving} a)). 

When pseudo-adiabatic averaging has a negligible influence, i.e. in ref.~\cite{Proux2015} $\overline{ \delta E^2 } /\overline{ E^2 } \simeq 3 \% \ll 1$, we finally recover $\overline {\g} (\tau) =\overline e^{-\Gamma_{\Las}|\tau|}e^{i\overline{\omega}_{\Las}\tau} \ \overline{\gtil} (\tau) \simeq e^{-\Gamma_{\Las}|\tau|}e^{i\overline{\omega}_{\Las}\tau}  \gtil (\tau)$ used in the case of monochromatic (non noisy) excitation.


\begin{figure}[h!]
\centering
\includegraphics[width=8cm]{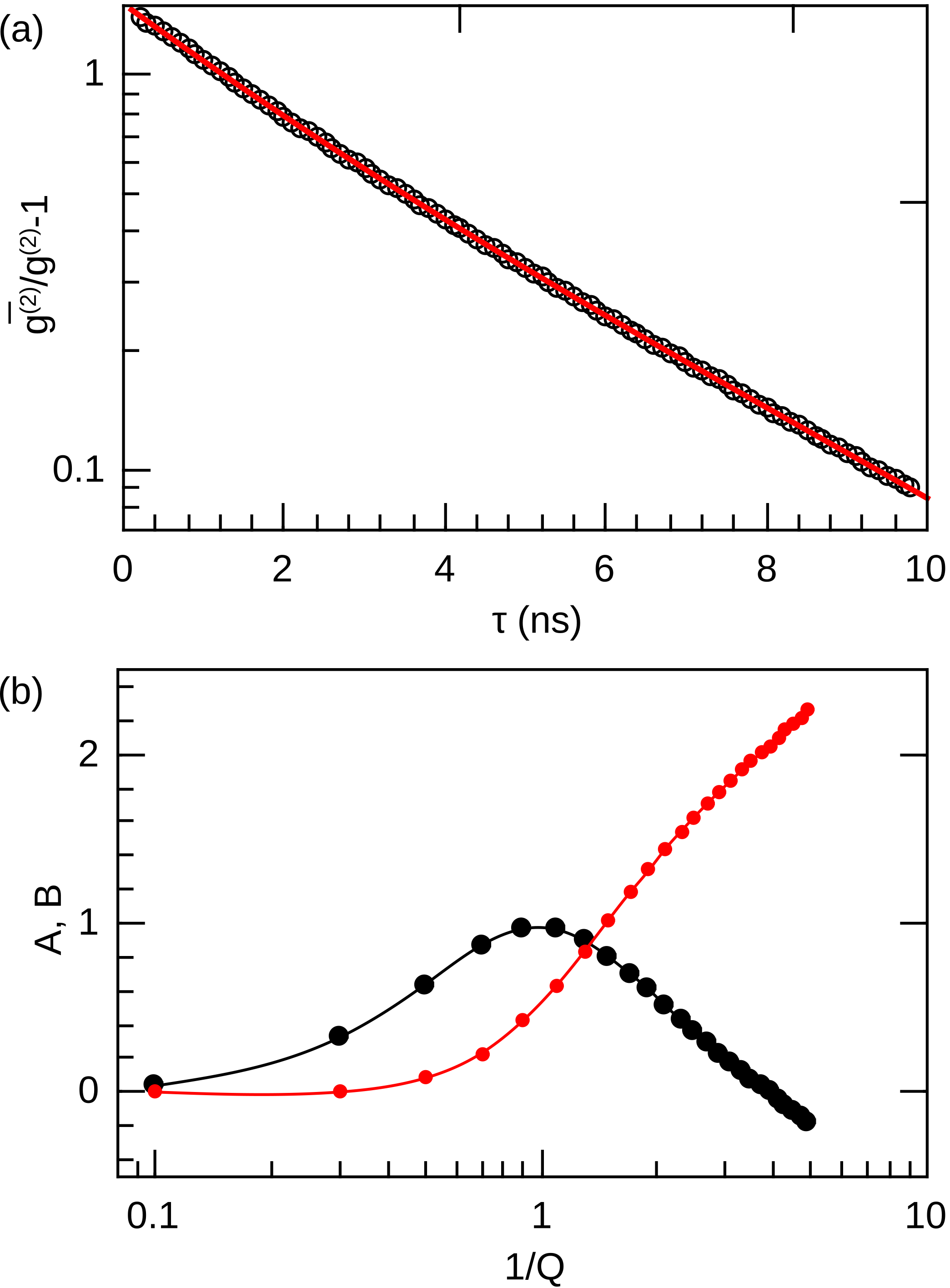}
\caption[graph]{
a) Reduced difference $\frac{\overline {\gt}(\tau)}{\gt (\tau)}-1$ between the pseudo-adiabatic simulation $\overline {\gt}(\tau)$ and the fluctuation-free $\gt (\tau)$ for the case considered in fig.~\ref{fig:Weak_driving} b) (thick black line). 
The solid red line indicates the fit result using the model proposed in eq.~(\ref{eq:A_q_B_q}) with $\tau_C= 4 \si{ns}$, $Q=1$, $A=1.$ and $B=0.5$.
b) Coefficients A (black) and B (red) as a function of $1/Q$ obtained from fitting numerical integrations of post-averaged Bloch-Liouville equation for $Q=[0.1;5]$. Thin plain lines are guide for the eye.
}
\label{fig:Aq_Bq_g2}
\end{figure}

For $\gt$, pseudo-adiabatic averaging is more complex and Taylor expansion has to be conducted up to fourth order. The result takes the form

\begin{equation}
\overline {\gt_{E, \Delta \omega } } (\tau) \simeq   \left( 1 + A(Q) e^{-\tau/\tau_C} + B(Q) e^{-2\tau/\tau_C} \right) \gttil (\tau),
\label{eq:A_q_B_q}
\end{equation}
where $A$ and $B$ are dimensionless coefficients function of $Q$. We recognize that the driving field fluctuations induce an extra bunching at short times reminiscent from the classical bunching that is observed in the noisy excitation second-order auto-correlation function as is observed on fig.~\ref{fig:Weak_driving} b). 
Figure~\ref{fig:Aq_Bq_g2} illustrates how $A(Q)$ and $B(Q)$ are obtained by fitting numerical simulations (such as fig.~\ref{fig:Weak_driving} b) ). A characteristic fit is represented on fig.~\ref{fig:Aq_Bq_g2} a). The result of a series of fits is represented on fig.~\ref{fig:Aq_Bq_g2} b).

\begin{figure}[h!]
\centering
\includegraphics[width=8cm]{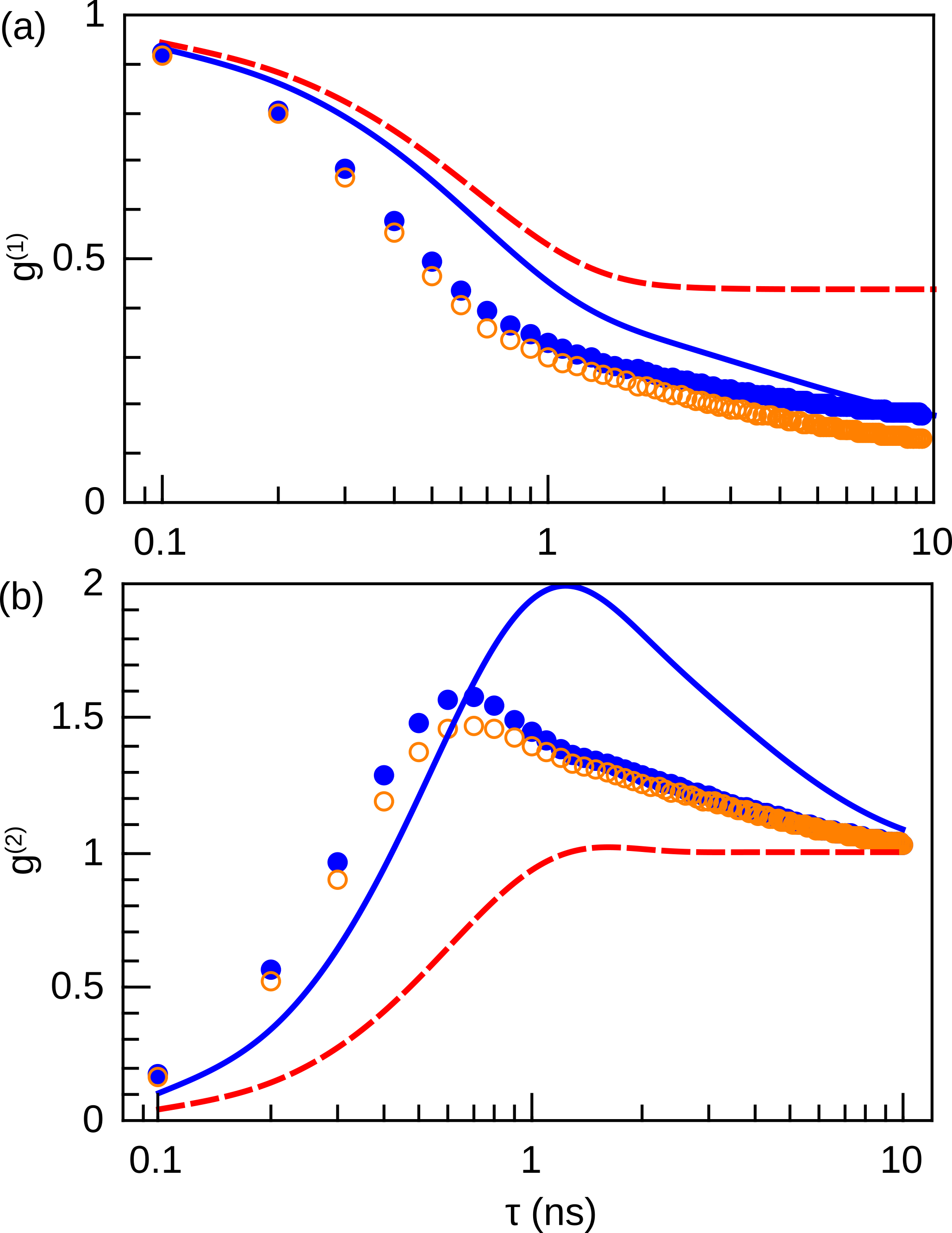}
\caption[graph]{ 
$\overline \gtil$ (a) and $\overline \gt$ (b) in the strong driving limit obtained by numerical integration of the Liouville equations taking into account pseudo-adiabatic averaging. The simulations are represented by symbols and are computed for a resonant excitation and for $\tau_C= 4 \si{ns}$, $T_1= 0.34 \si{ns}$ , $T_2= 0.5 \si{ns}$, $\overline \Omega=2. \si{rad/ns}$, and consequently a saturation parameter $s=0.68$ (strong driving). In both cases, $\sqrt{\overline{\delta \Omega^2}} = \sqrt{\overline{\delta \omega^2}}= 2. \si{rad/ns}$ which corresponds to $Q=1$. The rotating frame blurring is not included in the calculation. Both $\overline \gtil$ and $\overline \gt$ simulations have been done with a correlation factor between driving field amplitude and energy fluctuations $\epsilon=0$ (circles) and $0.8$ (dots). 
The fluctuationless theories are represented as dashed lines and the weak driving limit theory as thick lines. Both fail at describing the simulations when fluctuations are large. 
}
\label{fig:Strong_driving}
\end{figure}

Finally, figure~\ref{fig:Strong_driving} gives an example of numerically computed $\overline \gtil$ and $\overline \gt$ in the strong driving limit ($s\sim 1$), i.e., when inelastic scattering starts to dominate.
As expected, linear response theory does not correctly capture the correction due to the pseudo-adiabatic averaging as can be seen on figs.~\ref{fig:Strong_driving} a) and b) for $\g$ and $\gt$, respectively.
Moreover, as opposed to the weak driving case, the fluctuation correlation factor $\epsilon$ has a visible consequence on the pseudo-adiabatic result in contrast with the weak driving regime. 
However, despite smaller correlation functions than predicted by the linear response theory, we remark that qualitative features are partially preserved such as relevant timescales and extra bunching in $\gt$ at small delays. 

\section{Application to a Hong-Ou-Mandel interferometer}
\label{s:propagation}

\subsection{Photons indistinguishability}

The Hong-Ou-Mandel interferometer allows measuring  the second-order intensity cross-correlation function $\gtX$. 
Fig. 5 shows the principle of the basic Hong-Ou-Mandel interferometer we consider first. The fields emitted by two independent sources are sent at the two inputs of a 50:50 beamsplitter. Two detectors measure the intensity of light at the two outputs of the beamsplitter. If the response time of these detectors is faster than the mutual coherence time of the fields, a fourth order (or two-photon) interference effect can be detected. To characterize the interference, the cross-correlation function $\gtX$ of the intensity measured by the two detectors is calculated. This measurement is based on the idea that if the detectors both detect a photon, it will bring a peak in the correlation function. In the case of single photons, this can happen only if the photons are distinguishable. Therefore, for a single photon source, a non-zero $\gtX$ imply that detected photons are distinguishable to a certain degree. In practice, a dip in the $\gtX$ measurement is the signature of indistinguishable photons.

It is also possible to measure $\gtX$ using only one source: A first beamsplitter divides the emitted field  into two arms and a second beamsplitter recombines the field. If the path difference between the two arms is longer than the field coherence length, the two fields incoming on the second beamsplitter are completely independent.

\begin{figure}[h!]
\centering
\includegraphics[width=8cm]{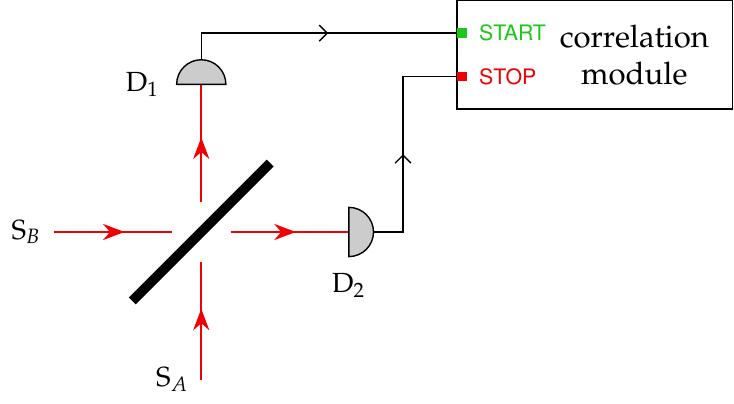}
\caption[graph]{
The cross-correlation experiment to characterize the indistinguishability of photon emission of two identical sources. The HOM uses only one source at consecutive times using an extra beamsplitter and an assymetric delay on the two arms. 
}
\label{fig:HOM_principle}
\end{figure}

Up to now, we only considered our system to be two photons interfering on a beamsplitter. We have implied an exact and controlled time of emission of the photons, or at least a control of the delay between subsequent photons, similarly to what would happen with a two-level system excited by a mode-locked pulsed laser\cite{Santori2002}. But if the emitter emits a continuous wave (cw), the detection time of the photons and thus the delay between these detections is random. This means that instead of being able to probe exact photon delays using accurately controlled and ultra-short excitation pulses, the experiment needs to time the photon arrivals accurately enough to resolve the change in intensity correlation due to two-photon interference as a function of the detection delay. This also means that if the distinguishability of the photons only arises from the driving energy fluctuations – as is most often the case – it just decreases the mutual coherence time of the two fields and therefore only reduces the width of the indistinguishability dip, and not its amplitude. The ability of the detection system to resolve this dip is what governs its amplitude, which then cannot be used to characterize indistinguishability in a cw HOM experiment. Therefore, a new figure is required to assess indistinguishability in a cw experiment.



\subsection{The coalescence time window}

The choice of a figure of merit is dictated by several key considerations, potentially contradictory:
\begin{enumerate}
\item It has to be a meaningful quantity, preferably unrelated to a precise modelization. In this case, it should be interpretable in terms of conditional probabilities;
\item It has to be independent of a precise measurement apparatus. (SPD response, interferometer alignement)
\item It has to allow quantitative comparisons between various single-photon sources, the fluctuating driving source being considered part of the single-photon source itself.
\end{enumerate}

An interpretation in terms of conditional probabilities of photon emission is possible with a two-level system only if this source has been characterized as a single photon source, i.e., in a previous HBT experiment, the $\Gt$ drops to zero at null delay. Supposing this is true, the new figure of merit is built as follow:

\begin{enumerate}
\item The two-photon component response is measured as $G^{(2\text{X})}_\perp - G^{(2\text{X})}_\parallel$, where $G^{(2\text{X})}_\perp$ is the intensity correlation function measured at the output of the HOM interferometer when the two arms are cross-polarised, while $G^{(2\text{X})}_\parallel$ is the equivalent when the polarisation of the arms is parallel.

\item The two-photon response at delay $\tau$ divided by the response with orthogonal polarizations, $\GtX_\perp$, defines the visibility $V_{\mathrm{HOM}}(\tau)$ and can be interpreted as the conditional probability of having two \textbf{indistinguishable} photons separated by a delay $\tau$ in the interferometer knowing that two single photons have been emitted.


\item From this last quantity, we can extract the theoretical CTW which is obtained as the average duration of the visibility $V_{\mathrm{HOM}}(\tau)$, i.e. $\int V_{\mathrm{HOM}}(\tau) \mathrm{d} \tau$. The CTW is independent of the SPD response time provided this response time is short compared to CTW.

\end{enumerate}

In this case, particularly relevant with nowadays lasers, CTW can be interpreted as the characteristic time over which the photons are considered indistinguishable. 

The CTW corresponds to the area under the curve of two-photon interference visibility. This means that the widening of the visibility curve due to the convolution of the measured $G^{(2\text{X})}$ by the detectors will have a small effect on the value of the CTW as long as it remains long compared to the response time of the detectors. Note that the CTW is well defined for fast detectors as opposed to what has been interpreted in early litterature, e.g. in \cite{kalliakos2016enhanced}. 

Under pulsed excitation, the two-photon interference visibility at zero delay of a two-level system at the low power limit is equal to $T_2/2T_1$, which characterizes how far the system is from the radiative limit. Hence a decreasing $T_2/2T_1$ lowers the value of the visibility at zero delay. On the contrary, under cw excitation, the value of the visibility at zero delay is always zero when accounting for the time response of the detectors. 

In contrast, under cw excitation, when $T_2/2T_1$ equals 1, the CTW is equal to the sum of the laser coherence time and the residual area due to the single photon dynamics – usually equal to $T_1$. A decreasing $T_2/2T_1$ lowers the CTW and when $T_2/2T_1$ is zero, the CTW is zero as well. The imperfections of the measurement (in particular interference visibility) also result in a reduction of the CTW by decreasing the visibility. However, these defects can be accounted for and CTW can be used to compare emitters in terms of photon indistinguishability.

\subsection{Expression of the correlation functions for the HOM experiment}
\label{ssec:Expr_HOM}

In this section, we describe the HOM interferometer response function. 
At this stage, we assume that the physical source of photons is a point-like electric dipole. 

\begin{figure}[h!]
\centering
\includegraphics[width=8cm]{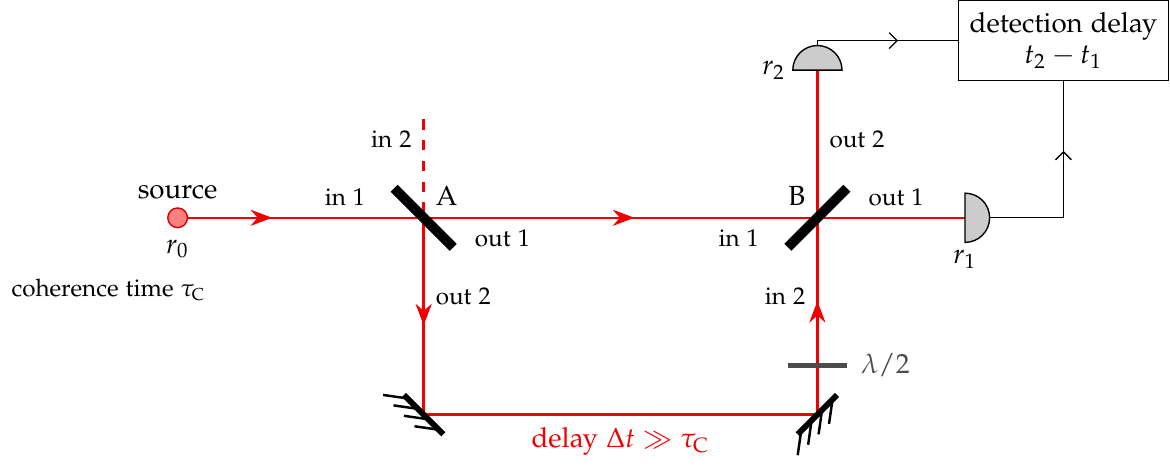}
\caption[HOM scheme]{Scheme of the Hong-Ou-Mandel interferometer. Beam-splitter A separates the field scattered by the two-level system. A delay $\Delta t$ is set between the two arms. Time-delayed fields beats on beam-splitter B and the time-correlation between the outcoming fields are characterized through the time correlation of two identical single-photon time-resolved detectors.}
\label{fig:notations}
\end{figure}

The HOM setup considered is represented on Fig.~\ref{fig:notations}. With this geometry, the electromagnetic field radiated by the source is split by a first beam splitter A and recombined on a second one B, the propagation delay between the two arms being fixed. Similarly to a HBT interferometer, the correlation function of the interfering field on the second beam splitter is measured.


Therefore, we are interested in the probability density function (PDF) of joint-photodetection
\begin{equation}
w(r_{1},t_{1},r_{2},t_{2})=
\sum_{f}\left|\langle f|TE^{(+)}(r_{2},t_{2})E^{(+)}(r_{1},t_{1})|i\rangle \right|^{2}
\label{e:MasseEtats1}
\end{equation}

\noindent where $|i\rangle$ and $|f\rangle$ are the initial and final states of the electric field, respectively. $E^{(+)}(r,t)$ and $E^{(-)}(r,t)$ are the positive and negative energy electric field operators at position $r$ and time $t$, respectively, with $r_{1,2}$ standing for the
positions of the two photodetectors and $t_{1,2}$ for the detection times. $T$ is the time-ordering operator. From this initial PDF expression we obtain the equivalent density-matrix expression~\cite{Scully1997}:
\begin{multline}
w(r_{1},t_{1},r_{2},t_{2})=\\
\Tr\big(E^{(-)}(r_{1},t_{1})E^{(-)}(r_{2},t_{2})T^{\dagger}TE^{(+)}(r_{2},t_{2})E^{(+)}(r_{1},t_{1})\hat{\rho}\big)
\label{e:MasseEtats2}
\end{multline}
where $\Tr$ stands for the trace operator and $\hat{\rho}$ is the density matrix describing the state of the complete electric field. The electric field at point $r_{1,2}$ and time $t_{1,2}$ is a function of the field at point $r_{0}$ (the position of the small emitter) and former times.

The notations we will use to introduce these correlation functions of the e.m. field are introduced on fig.~\ref{fig:notations}: The source, and detectors D1 and D2 are positioned at $r_0$, $r_1$ and $r_2$, respectively. The beam splitter A (B) has transmission and reflection coefficients $t_\mathrm{A}$ ($t_\mathrm{B}$) and $r_\mathrm{A}$ ($r_\mathrm{B}$).
The input and output ports are labeled $\mathrm{in\,1}$ ($\mathrm{in\,2}$) and $\mathrm{out\,1}$ ($\mathrm{out\,2}$), respectively, and their positions are, e.g. $r^{\mathrm{B}}_\mathrm{in\,2}$ for the input port 2 of beam splitter B. Propagation delays are indicated using the spatial reference of the path, i.e. $t_\mathrm{AB}$ for the delay between splitters A and B on the short arm of the interferometer.

We account for the polarization of light in the interferometer which can be controlled on the long arm through the use of a half wave-plate. The light on the short arm is assumed to be linearly polarized and define the polarization $\vec e_x$, whereas the light polarization on the long arm is polarized along $\cos\phi \vec e_x + \sin\phi \vec e_y$.

The following equations are the propagation relations of the e.m. fields in the various elements of the HOM interferometer:

\begin{equation}
\begin{split}
\vEp(r_1, t_1) &= \vEp(r^{\mathrm{B}}_\mathrm{out\,1}, t_1 - t_\mathrm{B1})\\
\vEp(r_2, t_2) &= \vEp(r^{\mathrm{B}}_\mathrm{out\,2}, t_2 - t_\mathrm{B2})
\end{split}
\end{equation}

Mixing at beam splitter B: 
\begin{equation}
\begin{split}
\begin{pmatrix}
\vEp(r^{\mathrm{B}}_\mathrm{out\,1}, t)\\
\vEp(r^{\mathrm{B}}_\mathrm{out\,2}, t)
\end{pmatrix}
&=
\begin{bmatrix}
t_\mathrm{B} & \iim r_\mathrm{B}\\
\iim r_\mathrm{B} & t_\mathrm{B}
\end{bmatrix}
\begin{pmatrix}
\vEp(r^{\mathrm{B}}_\mathrm{in\,1}, t)\\
\vEp(r^{\mathrm{B}}_\mathrm{in\,2}, t)
\end{pmatrix}
\end{split}
\end{equation}

Propagation in the two arms between beam splitters A and B:
\begin{equation}
\begin{split}
\vEp(r^{\mathrm{B}}_\mathrm{in\,1}, t) &= \Ep(r^{\mathrm{A}}_\mathrm{out\,1}, t - t_\mathrm{AB})\vec e_x\\
\vEp(r^{\mathrm{B}}_\mathrm{in\,2}, t) &= \Ep(r^{\mathrm{A}}_\mathrm{out\,2}, t - t_\mathrm{AB}-\upDelta t) \\
					& 		(\cos\phi\, \vec e_x + \sin \phi\, \vec e_y)
\end{split}
\end{equation}

Mixing at beam splitter A: 
\begin{equation}
\begin{split}
\begin{pmatrix}
\vEp(r^{\mathrm{A}}_\mathrm{out\,1}, t)\\
\vEp(r^{\mathrm{A}}_\mathrm{out\,2}, t)
\end{pmatrix}
&=
\begin{bmatrix}
t_\mathrm{A} & \iim r_\mathrm{A}\\
\iim r_\mathrm{A} & t_\mathrm{A}
\end{bmatrix}
\begin{pmatrix}
\vEp(r^{\mathrm{A}}_\mathrm{in\,1}, t)\\
\vEp(r^{\mathrm{A}}_\mathrm{in\,2}, t)
\end{pmatrix}
\end{split}
\end{equation}

Input fields:
\begin{equation}
\begin{split}
\vEp(r^{\mathrm{A}}_\mathrm{in\,1}, t) &= \Ep(r_0, t - t_\mathrm{0A})\vec e_x\\
\vEp(r^{\mathrm{A}}_\mathrm{in\,2}, t) &= \vec 0
\end{split}
\end{equation}

Combining the latter relations, we obtain
\begin{subequations}\label{eq:E1-E2}
\begin{align}
\vEp (r_1, t_1) &= t_\mathrm{A} t_\mathrm{B} \Ep (r_0, t)\vec e_x - \nonumber \\
		 r_\mathrm{A} r_\mathrm{B} &\Ep (r_0, t-\Delta t) (\cos\phi\, \vec e_x + \sin \phi\, \vec e_y)\\
\vEp (r_2, t_2) &= \iim t_\mathrm{A} r_\mathrm{B} \Ep (r_0, t+\tau)\vec e_x + \nonumber \\
		\iim r_\mathrm{A} t_\mathrm{B} &\Ep (r_0, t+\tau-\Delta t)  (\cos\phi\, \vec e_x + \sin \phi\, \vec e_y)
\end{align}
\end{subequations}
where we have defined
\begin{subequations}
\label{eq:def-t-tau}
\begin{align}
t &= t_1 - t_\mathrm{B1} - t_\mathrm{AB} - t_\mathrm{0A}\\
t+\tau &= t_2 - t_\mathrm{B2} - t_\mathrm{AB} - t_\mathrm{0A}.
\end{align}
\end{subequations}

Assuming that the emitter is weakly coupled to the electric field and only emits through electric dipolar radiative transitions,
then $E^{(\pm)}(r_{0},t_{0})=\alpha S^{\mp}(t_{0})$, where $S$ is the total dipole operator of the emitter transitions, and $\alpha$ is a coefficient determined by both setup and emitter properties and proportional to oscillator strength of the two-level system transition of interest and collection efficiency of the interferometer \cite{Cohen1988}. 

By neglecting the thermal e.m. field at input port 2 of beam splitter A, we obtain the following expression:
\widetext
\begin{equation}
w(r_{1},t_{1},r_{2},t_{2})=
\sum_{(\varepsilon_{1},\varepsilon_{2},\varepsilon_{3},\varepsilon_{4})\in\{0,1\}}A_{\varepsilon_{1},\varepsilon_{2},\varepsilon_{3},\varepsilon_{4}}\left\langle {S_+}(t-\varepsilon_{1}\Delta t){S_+}(t+\tau-\varepsilon_{2}\Delta t)T^{\dagger}T{S_-}(t+\tau-\varepsilon_{3}\Delta t){S_-}(t-\varepsilon_{4}\Delta t)\right\rangle
\label{e:HOMGalExpr}
\end{equation}
\endwidetext

\noindent where $T^\dagger$ is the backward time-ordering operator acting on the left part of the equation. These
time-ordering operators guarantee that causality is verified in the measurement process. Values $A_{\varepsilon_{1},\varepsilon_{2},\varepsilon_{3},\varepsilon_{4}}$ are given in table~\ref{t:HOMCOeffs}.

\begin{table*}[!ht]
\begin{tabular}{|c|c|c|c|}
\hline
\textrm{term} & \textrm{value} & \textrm{term} & \textrm{value} \\
\hline
0000 & $|\ta|^4 |\tb|^2 |\rb|^2$ & 1000 & $-|\ta|^2|\rb|^2 \ta^* \tb^* \ra\rb\,\cos\phi$ \\
0001 & $-|\ta|^2|\rb|^2 \ta \tb \ra^* \rb^* \,\cos\phi$ & 1001 & $|\ta|^2 |\ra|^2 |\rb|^4$ \\
0010 & $|\ta|^2|\tb|^2 \ta^* \tb \ra\rb^*\,\cos\phi$ & 1010 & $-|\ta|^2|\ra|^2 \tb^{2} \rb^{*2}\,\cos^2\phi$ \\
0011 & $-|\tb|^2|\rb|^2 \ta^{*2} \ra^2\,\cos^2\phi$ & 1011 & $|\ra|^2|\rb|^2 \ta^* \tb \ra\rb^*\,\cos\phi$ \\
0100 & $|\ta|^2|\tb|^2 \ta \tb^* \ra^* \rb\,\cos\phi$ & 1100 & $-|\tb|^2|\rb|^2 \ta^{2} \ra^{*2}\,\cos^2\phi$ \\
0101 & $-|\ta|^2|\ra|^2 \tb^{*2} \rb^2\,\cos^2\phi$ & 1101 & $|\ra|^2|\rb|^2 \ta \tb^* \ra^* \rb\,\cos\phi$ \\
0110 & $|\ta|^2 |\tb|^4 |\ra|^2$ & 1110 & $-|\tb|^2|\ra|^2 \ta \tb \ra^* \rb^* \,\cos\phi$ \\
0111 & $-|\tb|^2|\ra|^2 \ta^* \tb^* \ra\rb\,\cos\phi$ & 1111 & $|\tb|^2 |\ra|^4 |\rb|^2$ \\
\hline
\end{tabular}
\caption[HOM weights]{HOM weights. The terms are identified by their indexes $\varepsilon_{1},\varepsilon_{2},\varepsilon_{3},\varepsilon_{4}$.}
\label{t:HOMCOeffs}
\end{table*}

In the case of a noisy excitation source, the various terms in the Hong-Ou-Mandel expression (\ref{e:HOMGalExpr}) and the corresponding phase accumulation terms are listed in Table \ref{table:hom} which  reads as follows : let us consider term number 1, it corresponds to the following timings $\{0,\tau,\tau,0\}$ hence its formal expression is $\left\langle {S_+}(t){S_+}(t+\tau)T^{\dagger}T{S_-}(t+\tau){S_-}(t)\right\rangle$ and its prefactor is $A_{0,0,0,0}$ which value $ |\tb|^2 |\rb|^2 (|\ta|^4 + |\ra|^4)$ is obtained using the relations (\ref{e:MasseEtats2}) to (\ref{eq:def-t-tau}). 
We recognize that term 1 is simply the intensity auto-correlation function $\Gt (\tau)$ which is insensitive to the rotating frame transformation.

\begin{table*}[!ht]
\begin{tabular}{|c|c|c|c|c|c|}

\hline
\# & \textrm{factor} & \textrm{term} & \textrm{phase} & \textrm{averaged\ fluctuations} & \textrm{meaning}\\ \hline
1 & $ |\tb|^2 |\rb|^2 (|\ta|^4 + |\ra|^4)$ & \{0,$\tau$,$\tau$,0\} & 0 & 0 & $\Gt$($\tau$)\\ \hline
2 & $ |\ta|^2 |\ra|^2 |\tb|^4$ & $\{0,\tau-\Delta t,\tau-\Delta t,0\}$ & 0 & 0 & $\Gt(\tau-\Delta t)$\\ \hline
3 & $|\ta|^2 |\ra|^2 |\rb|^4$ & $\{-\Delta t,\tau,\tau,-\Delta t\}$ & 0 & 0 & $\Gt(\tau+\Delta t)$\\ \hline
4* & $|\ta|^2|\tb|^2 \ta^* \tb \ra\rb^*\,\cos\phi$ & $\{0,\tau,\tau-\Delta t,0\}$ & $\phi_{\tau\rightarrow\tau-\Delta t}$ & $e^{-\Gamma_{\Las}|\Delta t|}$ &\\ \hline
5* & $|\ra|^2|\rb|^2 \ta \tb^* \ra^* \rb\,\cos\phi$ & $\{0,\tau,\tau+\Delta t,0\}$ & $\phi_{\tau\rightarrow\tau+\Delta t}$ & $e^{-\Gamma_{\Las}|\Delta t|}$ &\\ \hline
6* & $-|\ta|^2|\rb|^2 \ta \tb \ra^* \rb^* \,\cos\phi$ & $\{0,\tau,\tau,-\Delta t\}$ & $\phi_{0\rightarrow-\Delta t}$ & $e^{-\Gamma_{\Las}|\Delta t|}$ &\\ \hline
7* & $-|\tb|^2|\rb|^2 \ta^{*2} \ra^2\,\cos^2\phi$ & $\{0,\tau,\tau-\Delta t,-\Delta t\}$ & $\phi_{0\rightarrow-\Delta t}+\phi_{\tau\rightarrow\tau-\Delta t}$ & $K$ &\\ \hline
8* & $-|\ta|^2|\ra|^2 \tb^{*2} \rb^2\,\cos^2\phi$ & $\{0,\tau-\Delta t,\tau,-\Delta t\}$ & $\phi_{0\rightarrow-\Delta t}+\phi_{\tau-\Delta t\rightarrow\tau}$ & $e^{-2\Gamma_{\Las}|\tau|}$ & $\textrm{2-photon\ interference}$\\ \hline
9* & $-|\tb|^2|\ra|^2 \ta^* \tb^* \ra\rb\,\cos\phi$ & $\{0,\tau-\Delta t,\tau-\Delta t,-\Delta t\}$ & $\phi_{0\rightarrow-\Delta t}$ & $e^{-\Gamma_{\Las}|\Delta t|}$ &\\
\hline
\end{tabular}
\caption{Terms of the Hong-Ou-Mandel interferometer response function (3 first columns), and the effect of the rotating frame blurring described in section \ref{ss:rfblurring} (columns 4 and 5). $K$ is $e^{-\Gamma_{\Las}(|\Delta t|+|\tau|)}$ if $|\tau|<|\Delta t|$, $e^{-2\Gamma_{\Las}|\Delta t|}$ otherwise.Terms 1,2,3 are the usual two-time intensity correlation functions whereas terms 4 to 9 are three- and four-time correlation functions. Among those last terms and for long delays between the two arms, only term 8 is significant under noisy resonant driving and it can be interpreted as the two-photon interference response of the interferometer, the other terms corresponding to one-photon interferences.}
\label{table:hom}
\end{table*}

A more interesting case is the term number 8 which is sensitive to the rotating frame transformation since 
\begin{multline*}
\left\langle {S_+}(t-){S_+}(t+\tau-\Delta t)T^{\dagger}T{S_-}(t+\tau){S_-}(t-\Delta t)\right\rangle =\\
e^{i \phi_{0\rightarrow-\Delta t}+\phi_{\tau-\Delta t\rightarrow\tau}} \times\\
\left\langle {\tilde{S}_{+}}(t-){\tilde{S}_{+}}(t+\tau-\Delta t)T^{\dagger}T{\tilde{S}_{-}}(t+\tau){\tilde{S}_{-}}(t-\Delta t)\right\rangle
\end{multline*}
We can then compute the average over phase fluctuations (rotating frame blurring) which corresponds to 
$$
\overline{ e^{i \phi_{0\rightarrow-\Delta t}+\phi_{\tau-\Delta t\rightarrow\tau}}} =
e^{-2\Gamma_{\Las}|\tau|}.
$$
Hence table \ref{table:hom} contains all the information necessary to compute the HOM interferometer response in the BPP case and it includes the rotating frame blurring effect. 

Three- and four-times correlation terms play an important role in the case of cw excitation when rotating frame blurring is absent. 
Among these, terms 4,5,6,7 and 9 correspond to one-photon interferences, i.e., the beating of the photonic field with itself for delays below the field coherence time. These one-photon interferences are analogous to classical interferences observed in a Mach-Zehnder interferometer. As they correspond to one-photon properties, they hinder the interpretation of the Hong-Ou-Mandel experiment as a way to measure the indistinguishability of photons which is provided by the two-photon properties. 

The use of a noisy laser source allows to decrease the laser coherence time $T_L=\Gamma_L^{-1}$. 
If the laser coherence time is sufficiently small compared to the
delay between the two arms, i.e.\ $\Delta t\gg \frac 1 {\Gamma_{\Las}}$, one-photon interference terms 4,5,6,7, and 9 are wiped out and only terms 1,2,3 and 8 (in table~\ref{table:hom}) remain and we can write the simplified unnormalized second-order intensity cross-correlation function (obtained for $\phi=\pi/2$): 
\begin{multline}
\GtX (\tau) = RT[(R^2 + T^2) \Gt(\tau)
			  + T^2 \Gt(\tau- \Delta t)+\\
			   R^2 \Gt(\tau+ \Delta t) ]
			  - 2R^2T^2\, e^{-2\Gamma_\Las |\tau|} \times \\
\big\langle {\tilde{S}_{+}} (0) {\tilde{S}_{+}} (\tau-\Delta t) {\tilde{S}_{-}} (\tau) {\tilde{S}_{-}} (-\Delta t) \big\rangle,
\label{e:HOMComplete}
\end{multline}

where we assume identical beam splitters A and B so that $R=|\ra|^2=|\rb|^2$ and $T=|\ta|^2=|\tb|^2$. If we further assume that $\Delta t\gg  T_{1},T_{2}$, the last term which is a four-time correlation function factorizes in a product of two-time correlation functions as follows:

$$\big\langle {\tilde{S}_{+}}(0){\tilde{S}_{-}}(\tau)\big\rangle \cdot\big\langle {\tilde{S}_{+}}(\tau-\Delta t){\tilde{S}_{-}}(-\Delta t)\big\rangle
=|\Gtil(\tau)|^{2}
$$

This simplification can be simply understood as the loss of memory of the state of the two-level system at time $\tau-\Delta t$ seen from $\tau$. 

The assumptions $\Delta t\gg \frac 1 {\Gamma_{\Las}}, T_{1},T_{2}$ are realized experimentally in ref. \cite{Proux2015} and the final expression is

\begin{multline}
\GtX (\tau) = RT[(R^2 + T^2) \Gt(\tau)
			  + T^2 \Gt(\tau- \Delta t)+\\
			   R^2 \Gt(\tau+ \Delta t) 
			  - 2RT\, e^{-2\Gamma_{\Las}|\tau|}|\Gtil(\tau)|^{2}],
\label{e:HOMCompletef}
\end{multline}

from which the normalized expression is obtained: 

\begin{equation}
\gtX (\tau) = \GtX (\tau) / [2RT (R^2+T^2) \overline{I}^2 ],
\label{e:HOMCompletef2}
\end{equation}
where $\overline{I^2}$ is the squared-field average intensity. Note that this response obtained under cw driving differs from the one under non-resonant driving\cite{patel2008postselective} which has been sometimes used in the non-noisy resonant driving regime, e.g. in \cite{Ates2009, weiler2013postselected, paudel2018generation}. 
The difference lies in the introduction of the $|\Gtil(\tau)|^{2}$ term which describes the influence of the driving resonant source. 



Expressions (\ref{e:HOMCompletef}) and (\ref{e:HOMCompletef2}) are only valid in the BPP regime. In the pseudo-adiabatic regime, it is necessary to further average over realizations. In this case, a special care must be devoted to the averaging $\overline{|\Gtil(\tau)|^{2}}$ which leads to $|\overline{\Gtil}(\tau)|^{2}$ since the HOM interferogram involves first-order correlation functions at two different times $0$ and $\Delta t$. 
By substituting this result into (\ref{e:HOMCompletef}), we obtain an effective relation in the noisy weak driving limit between the normalized correlation functions: 

\begin{multline}
\gtX (\tau) = W_2(\tau) \gt(\tau) + \\
\frac{ \alpha^2 W_2(\tau- \Delta t) \gt(\tau- \Delta t) + W_2(\tau+ \Delta t) \gt(\tau+ \Delta t)}{\alpha^2+1} - \\
 W_1(\tau) |\gtil(\tau)|^{2}
\label{eq:gtXaveraged}
\end{multline}

with

\begin{multline*}
W_1 (\tau) = \frac{\alpha \ e^{-2\Gamma_{\Las}|\tau|}}{1+\alpha^2} \frac{1}{(1+Q^{-2})^2} \left[ (1+Q^{-2} e^{-\tau/\tau_C} )^2 +		 \right. \\
    2 Q^{-2} e^{-\Delta t/\tau_C} (1+2 Q^{-2} e^{-\Delta t/\tau_C})+ \\
\left.    Q^{-2} ( e^{-|\Delta t+\tau|/\tau_C}+ e^{-|\Delta t-\tau|/\tau_C}) \right]
\end{multline*}

and

$$
W_2 (\tau) = 
\frac{1}{2} \left( 1+ A(Q) e^{-|\tau|/\tau_C} + B(Q) e^{-2|\tau|/\tau_C} \right)
$$

where $\alpha = R/T$. Expression (\ref{e:HOMCompletef2}) is recovered when fluctuation amplitudes are small ($Q^{-2} \rightarrow 0$). The new contributions describe the effect of the driving bunching on the HOM interferogram as observed on Fig.~\ref{fig:Weak_driving} and discussed in the following section.


\subsection{Application to the characterization of single photon indistinguishability}

Let us finally discuss the implications for the characterization of a single photon emitter. 
As seen in the previous section, a necessary requirement is that laser fluctuations are sufficient to wipe out one-photon intereferences and restore a meaning to the HOM interferogram. 
When this condition is met, it is possible to analyze the contributions of the fluctuations effects on the HOM interferogram. 

Before presenting the results, let us first recall that, as reported in section , $\g$ and $\gt$ correlation functions are independent of the correlation strength between the amplitude and phase fluctuations of the driving field. This observation is also true for the HOM response $\gtX$. This occurs because of the decoupling between the rotating frame blurring and pseudo-adiabatic averaging. It turns out to be a very fortunate fact since (i) the characterization of the amplitude-phase correlation is experimentally not trivial and (ii) the randomization of a laser source by a random driving is most likely to induce a large amplitude-phase correlation. This desirable property is not met above saturation, but this regime is less relevant for photon indistinguishability characterization. As a consequence, simulations reported in this section have been done using a driving field with decoupled amplitude and phase fluctuations, the coupled cases yielding identical results.

\begin{figure}[h!]
\centering
\includegraphics[width=8cm]{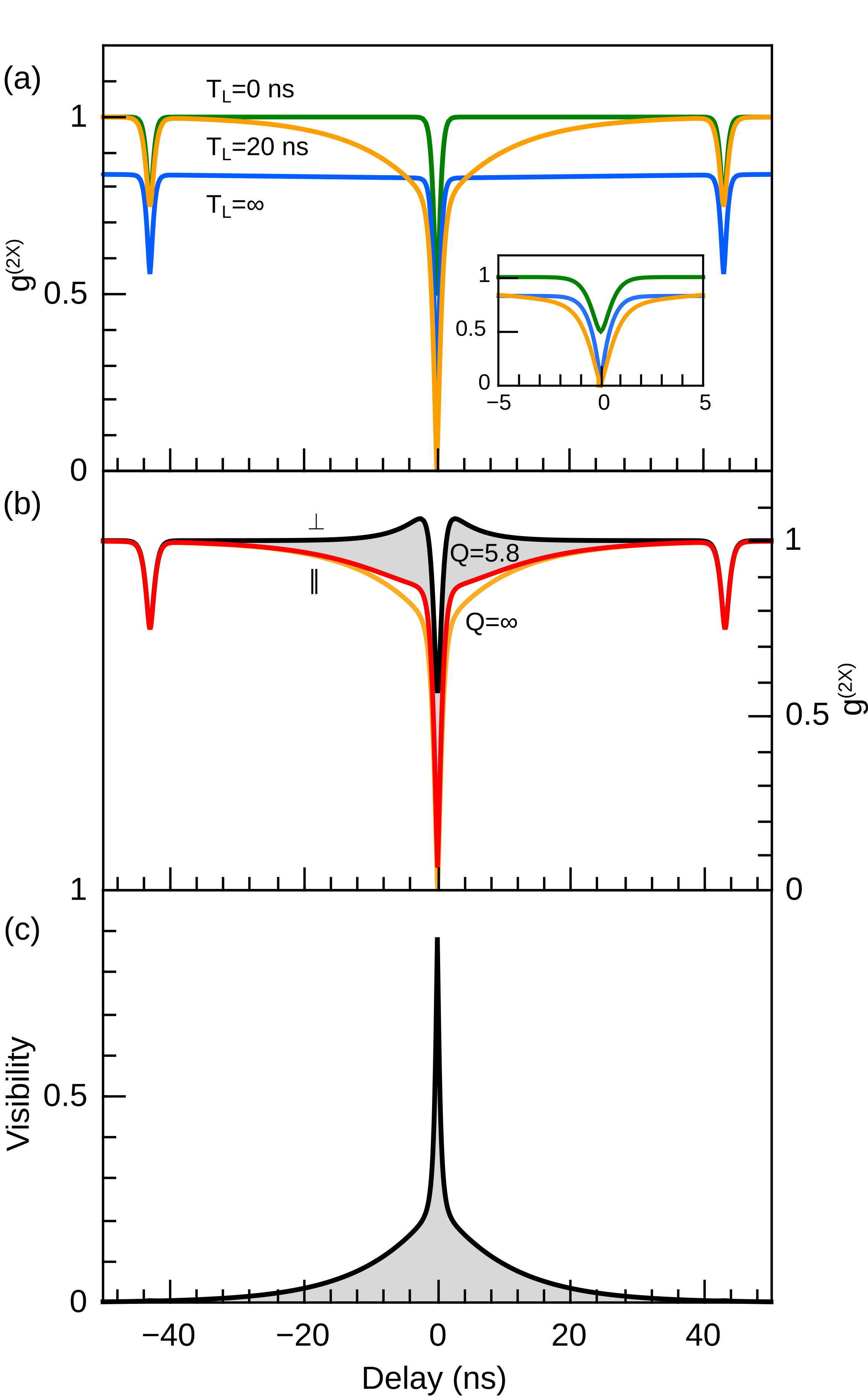}
\caption[graph]{Comparison of computed HOM interferograms obtained without and with noisy driving. 
(a) Effect of the driving energy fluctuations alone on HOM interferograms $\gtX_\parallel$, the rotating frame blurring effect is captured by the laser coherence time $T_L=\Gamma^{-1}_L={0,20,\infty}\ \si{ns}$ (green, orange and blue, resp.). Inset: zoom around zero delay.
(b) Effect of the driving field amplitude fluctuations $Q={1,\infty}$ (red and orange, resp.) for a fixed laser coherence time $T_L=20\si{ns}$. The reference cross-polarized interferogram $\gtX_\perp$ for $Q=1$ (black) is represented as well. 
(c) Visibility computed from panel (b) as $V(\tau)=|\gtX_\parallel (\tau)-\gtX_\perp (\tau)|/ \gtX_\perp (\tau)$
All simulations are done for a delay $\Delta t=43\si{ns}$, a field correlation time 4\si{ns}, a field amplitude  $\Omega = 0.1\si{rad/ns}$ corresponding to a saturation parameter $s=1.7\cdot 10^{-3}$, $T_1=0.34$\si{ns}, $T_2=0.5$\si{ns} and driving field amplitude fluctuation parameter $Q=5.8$ ($Q^{-2}=3\%$).
}
\label{fig:Visibility}
\end{figure}

Figure \ref{fig:Visibility} (a) represents the HOM response computed without and with laser phase fluctuations effects highlighting the contribution due to rotating frame blurring: In the absence of phase flucutations, $T_L \rightarrow \infty$ and is consequently larger than the interferometer delay $\Delta t$, the normalization of the HOM response is not trivial and the three-time correlators have a non-negligible contribution spoiling the interferogram interpretation. In the opposite limit where $T_L \rightarrow 0$, only intensity second-order correlation terms survive and the scattered single photons are completely distinguishable.

From the interferogram of panel (a), we obtain $\gtX(0)$. 
In all cases considered here, the simulations with balanced beam-splitters $\alpha = R/T=1$ yield to $\gtX(0)=0$. Indeed, from eq.~(\ref{eq:gtXaveraged}) we expect $\gtX(0)=\frac{1}{2}(\frac{\alpha-1}{\alpha+1})^2$. Even in the unbalanced beam-splitters case, $\gtX(0)$ is independent of the saturation parameter. 
This fully confirms that $\gtX(0)$ has no meaning both in the continuous excitation regime and the noisy resonant driving regime. 

The HOM response is also computed without and with driving field amplitude fluctuations effects (panel (b)) highlighting the bunching contribution due to adiabatic averaging at short times mostly observable on the crossed-polarized correlation function $\gtX$ at characteristic delays $\tau_C$. The comparison between the co- and crossed-polarized interferograms allows for the computation of visibility (panel (c)) from which the CTW is obtained. 

\begin{figure}[h!]
\centering
\includegraphics[width=8cm]{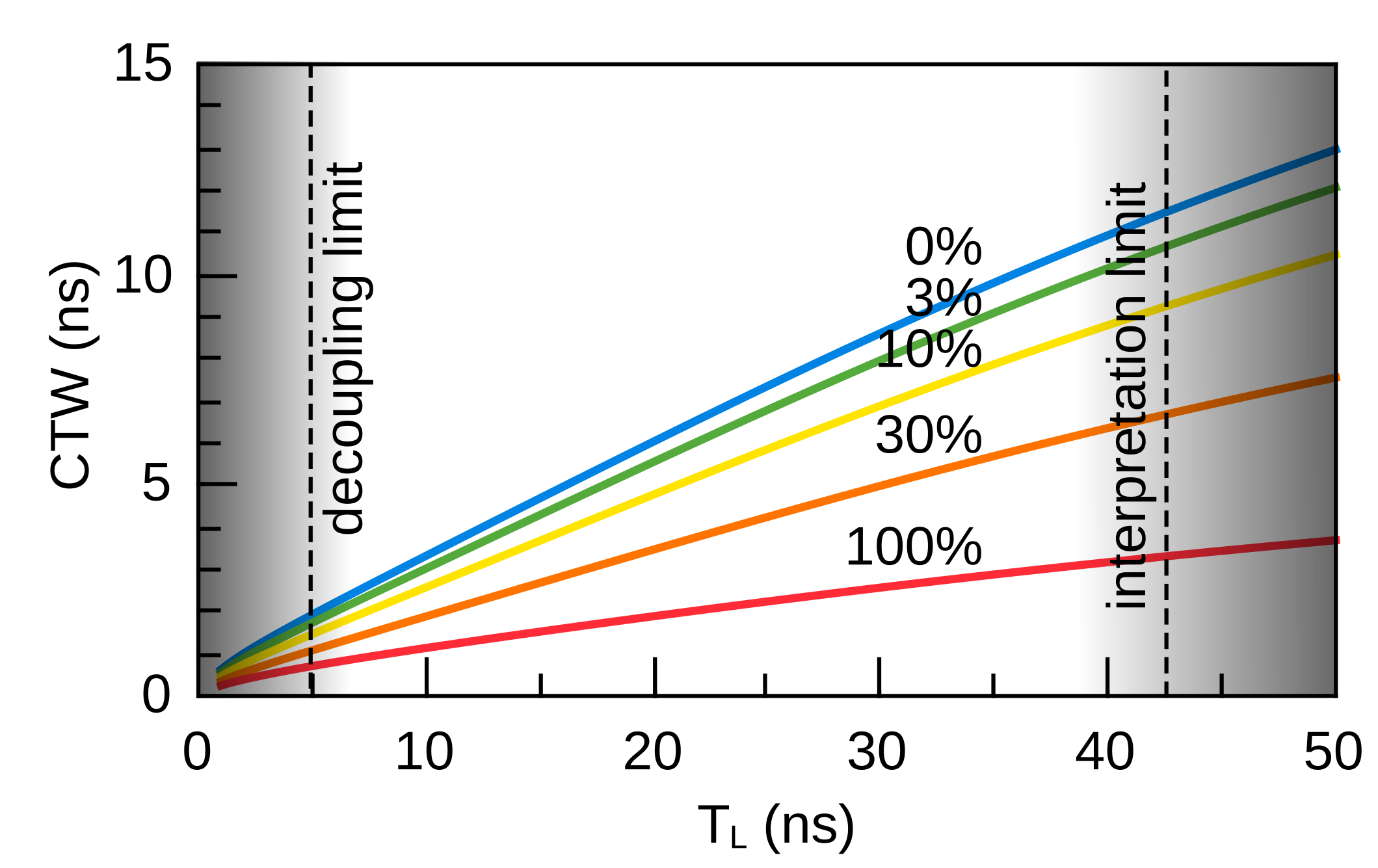}
\caption[graph]{CTW function of the laser coherence time $T_L$ for various diving field amplitude fluctuation strengths $Q^{-2}$ indicated on the figure. Simulation parameters are identical to fig.~\ref{fig:Visibility} lower panel. For $T_L \leq \tau_c$, the decoupling condition between rotating frame blurring and Bloch evolution is not fulfilled and the simulation result is not accurate. For $T_L \geq \Delta t$, the CTW can be computed but it cannot not be interpreted due to the single-photon self-interference terms as in the noiseless continuous excitation regime.}
\label{fig:g2XandCTWvssat}
\end{figure}

\begin{figure}[h!]
\centering
\includegraphics[width=8cm]{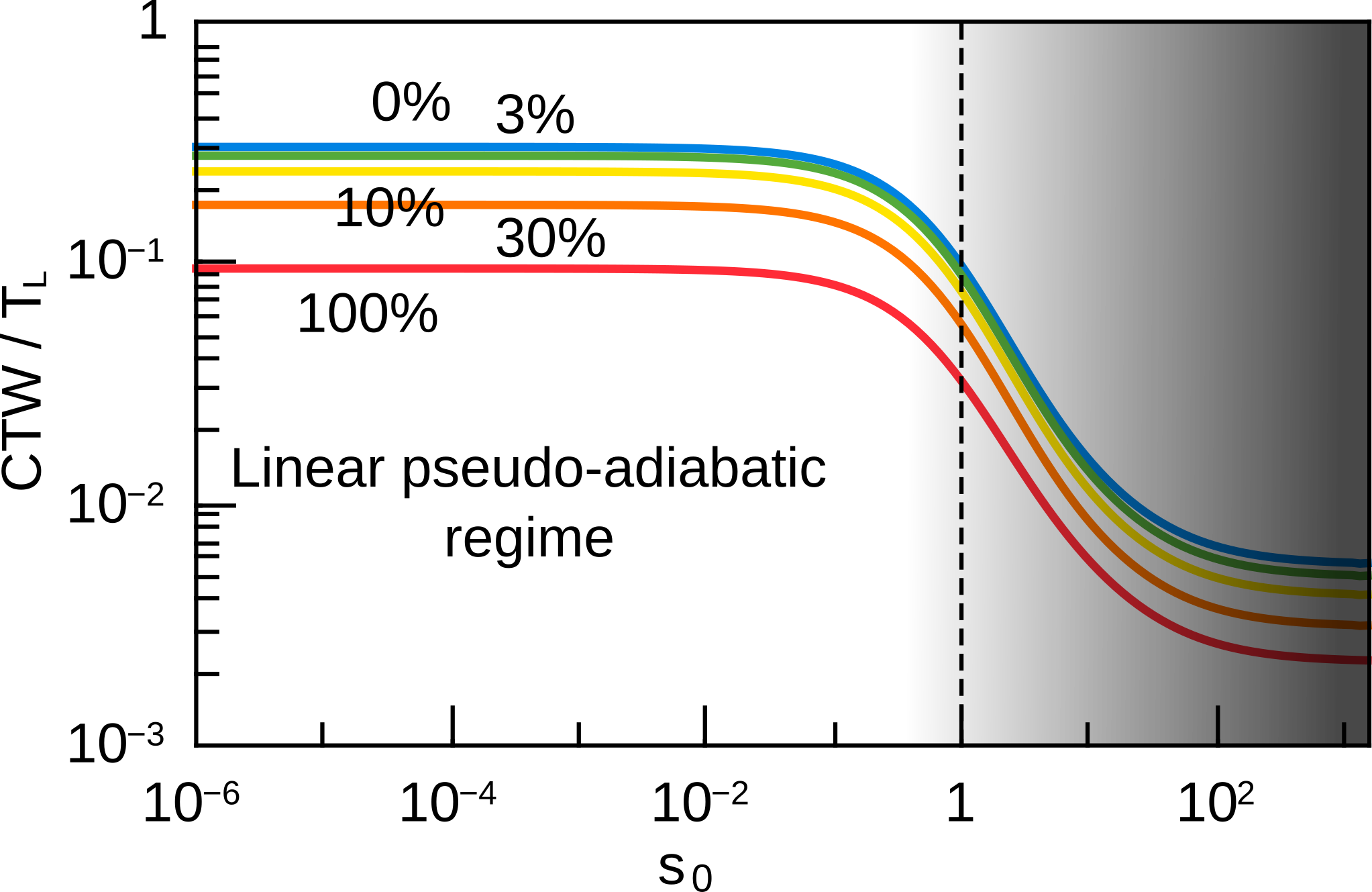}
\caption[graph]{CTW/$\mathrm{T}_L$ function of the driving amplitude expressed in terms of saturation parameter $s_0$ for various driving field amplitude fluctuation strengths (on the figure) and a fixed laser coherence time $T_L=20$~\si{ns}. Above saturation ($s_0\geq 1$), the linear approximation to the pseudo-adiabatic regime used in this paper is not valid anymore (although a limit theory can be done for small $Q^{-2}$.) The simulation parameters are the same as in fig.~\ref{fig:Visibility}. 
 }
\label{fig:g2XandCTWvsfluct}
\end{figure}

Figure \ref{fig:g2XandCTWvssat} represents the CTW when varying the laser coherence time $T_L$ at a fixed saturation $s_0=1.7\cdot 10^{-3}$. The limits of validity of the computation are represented on the graph and the interferometer delay $\Delta \tau$ is represented as reference. One can observe that the CTW and the laser coherence time are positively correlated, the CTW being always shorter than $T_L$. 
This latter feature is induced by the classical bunching of the driving that is transfered to the elastically scattered photons which destroys and single-photon indistinguishability characteristics.

Figure \ref{fig:g2XandCTWvsfluct} represents the CTW as a function of the saturation parameter. The CTW largely depends on the driving amplitude due to the change in the ratio of elastically/inelastically scattered photons. As expected the ratio $CTW/ T_L$ reduces with the driving field amplitude fluctuation strength $Q^{-2}$ due to bunching. In the experiment of ref~\cite{Proux2015} $Q^{-2}\sim 3$\% so that pseudo-adiabatic averaging brings only a small correction to the theoretical CTWs reported in this paper which were computed using eq. (\ref{e:HOMCompletef2}).

In general the CTW scales with the laser coherence time $T_L$, and is maximized when light is elasticaly scattered by the two-level system (below saturation) and for a lifetime limited two-level system ($T_2=2T_1$). Note that those requirements are qualitatively similar for the maximization of indistinguishability in the pulsed excitation regime.

\section{Conclusion}
\label{s:concl}

In this work we have explored the influence of a noisy source on the driving of single photon emitters and in particular on the statistics and indistinguishability of the emitted photons.
A wide range of behaviors is expected, but only a portion  of them, for which the dynamics is sufficiently well behaved, is simply interpretable and usable for the realization and characterization of antibunched light sources. In particular we have shown that the zero-delay intensity autocorrelation value $\gtX (0)$ used as figure of merit for photon indistinguishability in the pulsed excitation regime has no meaning in the continuous excitation regime. 
Consequently, we have introduced and justified an adequate figure of merit for photon indistinguishability in the continuous excitation regime: the coalescence time window. We have verified using numerical simulations that, in the relevant regimes for continuous quantum optics, the CTW indeed provides valuable information on single photon indistinguishability and we have shown how it is affected by the noisy driving source characteristics. It appeared that the maximization of the CTW in the noisy continuous excitation regime has similar requirements with the $\gtX(0)$ minimization in the pulsed excitation regime, namely that the two-level system should be linewidth-limited and operated in the elastic regime (low intensity). 
These results will allow classification and comparison of cw indistinguishable single photon sources for quantum optics and quantum cryptography.

\section*{ACKNOWLEDGEMENTS}
This work was financially supported by the French “Agence Nationale de la Recherche” (ANR-11-BS10-010) and “Direction Générale de l’Armement” (DGA).

\appendix
\section{Bloch-Purcell-Pound relaxation}
\label{a:BPP}

If correlation times of fluctuations are much faster than the two-level system dynamics in the rotating frame, the Bloch-Purcell-Pound (BPP) relaxation regime is reached.
We won't repeat the BPP derivation and its assumptions which are available in ref. \cite{Abragam1962} (chapter VIII).
In this appendix we provide explicit expressions for the BPP relaxation terms in the case of a two-level system in the ``non-viscous liquid'' case ($ \overline \Omega \tau_C\ll 1$) (we do not assume extreme narrowing here which is unphysical in the case of light emitters).
As a starting point, we consider the Hamiltonian $H(t) = H_0 + \overline H_1 + \delta H_1(t)$ and go in the rotating secular frame defined by $H_0 + \overline H_1$ so that the evolution equation reads

\begin{equation}
i \hbar \partial_t \tilde \rho(t) =[ \delta \tilde H_1(t) , \tilde \rho(t)].
\end{equation}

Integrating by successive approximations up to the second order leads to

\begin{equation}
 \partial_t \tilde \rho(t) = \frac{-1}{ \hbar^2}  \int_0^t \mathrm{d} t' [\delta \tilde H_1(t) , [\delta  \tilde H_1(t') , \tilde \rho(0)]].
\end{equation}

Using the usual series of assumption for Markovian decoherence~\cite{Abragam1962}, the following expression is obtained:

\begin{equation}
\partial_t \tilde \rho(t) =\frac{-1}{ \hbar^2}  \int_0^\infty \mathrm{d} \tau [\delta  \tilde H_1(t) , [\delta \tilde H_1(t-\tau) , \tilde \rho(t)]].
\end{equation}

Assuming $|H_0| \gg |\overline H_1|$, the fluctuating Hamiltonian reads~\cite{Abragam1962}:

\begin{equation}
\delta \tilde H_1(t) \simeq \hbar \Omega \frac{\delta E(t)}{\overline{ E } } \tilde S_x + \hbar \delta \omega(t) \tilde S_z.
\end{equation}

Consequently, we deduce the corresponding collapse operators for the Lindblad equation:

\begin{equation}
\left\{
\begin{array}{c}
L_1 = \sqrt{2\tau_C (\frac {\overline{ \Omega }^2}{\overline{ E }^2}  \overline{ \delta E^2 } - \frac {\overline{ \Omega }}{\overline{ E }} \overline{ \delta \omega \delta E })}   \tilde S_x\\
L_2 = \sqrt{2\tau_C (\overline{ \delta \omega^2 } - \frac {\overline{ \Omega }}{\overline{ E }} \overline{ \delta \omega \delta E } )}   \tilde S_z\\
L_3 = \sqrt{2\tau_C \frac {\overline{ \Omega }}{\overline{ E }} \overline{ \delta \omega \delta E } }    (\tilde S_x+ \tilde S_z)
\end{array} \right.
\end{equation}






In the ``non-viscous liquid'' case ($ \sqrt{\overline{ \delta \omega^2 } } \tau_C\ll 1$), extended and complex collapse operators can be found using a similar method by going into the doubly tilted rotating frame, see~\cite{Abragam1962, Tomita1958}. A remarkable property of the resulting collapse operators is their $\frac{\tau_C}{1+\Omega^2 \tau_C^2}$ dependence.

\section{Fluctuating frame averaging}
\label{a:fluctframe}

In this appendix, we show how to compute exactly the averaging over dephasing for the fluctuation characteristics defined in section~\ref{s:rotframe}.
It consists in averaging the dephasing term over phase distribution:
\begin{equation}
\overline{ e^{i \phi_{0\rightarrow t}} } = \int_{R} e^{i \phi} \mu (\phi,t|0,0)\  d\phi,
\end{equation}
where $\mu (\phi,t|0,0)$ is the conditional phase distribution.
If this distribution is assumed to be a normal law of mean 0 and variance $\phi_0^2(t)$, then integration leads to
\begin{equation}
\overline{ e^{i \phi_{0\rightarrow t}} } = e^{-\frac{\phi_0^2(t)}{2}}.
\end{equation}

The usual assumption (e.g., see ref.~\cite{Scully1997} (chapter 11.4)) consists in considering that the phase results from the accumulation of a white gaussian noise and consequently follows a Brownian trajectory. Consequently, the corresponding phase distribution at time $t$ is a normal law of mean $0$ and variance $\phi_0^2(t)= 2 \overline{\delta \omega^2} \tau_C t$. This concludes the computation of the averaging in this simple case which is
\begin{equation}
\overline{ e^{i \phi_{0\rightarrow t}} } = e^{- \Gamma_L t},
\end{equation}
with $\Gamma_L =\overline{\delta \omega^2} \tau_C$.

This approximate result is the useful one in most cases. However, it may be interesting to go beyond the Brownian motion approximation, first to investigate the validity of this approximation, secondly to obtain an exact result in the general case considered in section~\ref{s:fluctsource}.

This computation is possible in the case where two extra assumptions over the energy fluctuation statistics are made:
\begin{itemize}
\item the fluctuations result from a memoryless process;
\item the two-time correlation functions of the fluctuations maximizes Shannon entropy.
\end{itemize}
Those two assumptions are not very restrictive since they correspond to the case where fluctuations result from a low-pass filtered white gaussian noise which is close from the typical experimental realization (random or Johnson-Nyquist noise generator feeding a linear circuit).
It can be shown that the method used here is valid for any filtered white gaussian noise.

Assumptions over the fluctuations of $\delta \omega$ imply that it is governed by the following Langevin equation:
\begin{equation}
\frac{d \delta \omega}{ d t} = - \delta \omega/\tau_C + \eta(t),
\label{e:Lang1}
\end{equation}
where $\eta$ is a white gaussian noise of auto-correlation function $\overline{\eta (t) \eta (t')} =2 \overline{\delta \omega_0^2} \delta (t-t') /\tau_C$. Note that, from now on, the equilibrium variances are labeled $\overline{\delta \omega_0^2}$ to distinguish them from the 'out-of-equilibrium' variances $\overline{\delta \omega^2}$ involved in the Langevin equations. It can be easily shown that this results in the following statistics: $\overline {\delta \omega(t+\tau) \delta \omega(t)}=\overline{\delta \omega_0^2} e^{-t/\tau_C}$ and that $\delta \omega$'s PDF is a normal law of mean $0$ and variance $\overline{\delta \omega_0^2}$.
Phase accumulation $\phi$ is related to $\delta \omega$ through the equation:
\begin{equation}
\frac{d  \phi}{d t} =\delta \omega.
\label{e:Lang2}
\end{equation}
From eq.~(\ref{e:Lang1}) and (\ref{e:Lang2}) we deduce the following differential equations for the variance and covariance:
\begin{equation}
\frac{d  \overline{ \phi^2 }}{d t} = 2 \overline{ \phi \delta \omega  },
\label{e:Variance}
\end{equation}

\begin{equation}
\frac{d  \overline{ \phi \delta \omega  } }{d t} = -\frac 1 {\tau_C} \overline{ \phi \delta \omega  } + \overline{ \delta \omega^2  } .
\label{e:CoVariance}
\end{equation}

For which the solution is

\begin{equation}
\overline{ \phi^2 } (t) = 2 \overline{\delta \omega_0^2} {\tau_C} (t+\tau_C (e^{-\frac{t}{\tau_C}}-1)).
\label{e:GaussianSolution}
\end{equation}

At times longer than $\tau_C$, the phase accumulation of the Brownian motion is recovered while at shorter times than $\tau_C$ the phase variance evolves as $\overline{\delta \omega_0^2} t^2$ which can be understood as the ballistic behavior of phase accumulation at short times.

\section{Pseudo-adiabatic averaging}
\label{app:pseudoad}

To realize the pseudo-adiabatic integral, the joint distribution $p(\delta E_2,\delta \omega_2, t_2;\  \delta E_1,\delta \omega_1, t_1)$ is required. We will work in the reduced representation where time units are in $\tau_C$, and angular frequencies (field amplitude) units are in the corresponding equilibrium standard deviation of the equilibrium distribution, \textit{i.e.} the dimensonalized equations are recovered using the following substitutions $\delta \omega \rightarrow \delta \omega / \sqrt{\overline{\delta \omega_0^2}}$, $\delta E \rightarrow \delta E / \sqrt{ \overline{\delta E_0^2} } $, and $t \rightarrow t/\tau_C$.
Using the assumptions on the random variables $\delta E$ and $\delta \omega$ defined in section~\ref{s:deffluct}, we can write the corresponding Langevin equations: 


\begin{equation}
\left\{
\begin{array}{c}
\frac{\dd \delta \omega}{\dd t} = - \delta \omega  + \eta_1 (t)\\
\frac{\dd \delta E}{\dd t} = - \delta E + \eta_2 (t)
\end{array} \right.
\label{eq:Langevin}
\end{equation}

where $\eta_{1,2}(t)$ are two Langevin forces with the following correlation characteristics : 

$\overline{\eta_1 (t') \eta_1 (t)} = \overline{\eta_2 (t') \eta_2 (t)} = 2 \delta(t-t')$, $\overline{\eta_1(t') \eta_2 (t)} = \overline{\eta_2(t') \eta_1 (t)} = 2 \epsilon \delta(t-t')$, 
where $\epsilon$ is the correlation coefficient between energy and amplitude fluctuations. 
This is the unique set of Langevin equations given the constraints on the model provided in section~\ref{s:deffluct}.
Using Itô transform, one get the corresponding Fokker-Planck equation:

\begin{equation}
\begin{array}{c}
\frac{\partial p}{\partial t} = 
(\frac{\partial^2 }{\partial \delta \omega^2} + 
\frac{\partial^2 }{\partial \delta E^2} + 
2 \epsilon \frac{\partial^2 }{\partial \delta E \partial \delta \omega }) p +\\
 \frac{\partial ( \delta \omega p)}{\partial \delta \omega}+
\frac{\partial ( \delta E p)}{\partial \delta E}
\end{array}
\end{equation}

This partial differential equation has a simple generalized Gaussian solution fully characterized by its first and second order moments. Equations giving the dynamics of those moments are obtained using the Langevin equations~(\ref{eq:Langevin}):

\begin{equation}
\left\{
\begin{array}{c}
\frac{\dd \overline{\delta \omega}}{\dd t} = - \overline{\delta \omega}\\
\frac{\dd \delta \bar E}{\dd t} = - \delta \bar E  \\
\frac{\dd \overline{\delta \omega^2}}{\dd t} = - 2 \overline{\delta \omega^2} + 2 \overline{\eta_1 \delta \omega}\\
\frac{\dd \overline{\delta E^2}}{\dd t} = - 2 \overline{\delta E^2} + 2 \overline{\eta_2 \delta E}\\
\frac{\dd \overline{\delta E \delta \omega}}{\dd t} = - 2 \overline{\delta E \delta \omega} + \overline{\eta_2 \delta \omega} + \overline{\eta_1 \delta E}\\
\end{array} \right.
\label{eq:CoupledMoments}
\end{equation}


The source terms in the right-hand side second-moment equations (\ref{eq:CoupledMoments}) are respectively $2 \overline{\eta_1 \delta \omega}$, $2 \overline{\eta_2 \delta E}$, $\overline{\eta_2 \delta \omega} + \overline{\eta_1 \delta E}$. Their values are obtained using the following considerations: 

$$ 
\overline{\eta_1 \delta \omega}=
\overline{\eta_1 (t) \delta \omega (t)} = 
\frac{1}{2} (\overline{\eta_1 (t) \delta \omega (t^+)} + \overline{\eta_1 (t) \delta \omega (t^-)})
$$

But $\overline{\eta_1 (t) \delta \omega (t^-)} =0$ since, due to causality, there is no correlation between $\eta_1 (t)$ and $\delta \omega (t^-)$. 

Now, 
$$
\overline{\eta_1 (t) \delta \omega (t^+)} = \overline{\eta_1 (t) \delta \omega (t^-)} + \int_{t^-}^{t^+} \dd t' \frac{\partial}{\partial t'} \overline{\eta_1 (t) \delta \omega (t')}.
$$

Using the Langevin equations and the Langevin force correlation function, we find that the integral in the right hand-term is simply given by the integral of a Dirac function so that $\overline{\eta_1 (t) \delta \omega (t^+)} = 2$. 
Finally we then have $\overline{\eta_1 \delta \omega}= 1$. 
Identically, $\overline{\eta_2 \delta E}= 1$, $\overline{\eta_2 \delta \omega} = \epsilon$, and $\overline{\eta_1 \delta E}=\epsilon$. 

The set of linear differential equations~(\ref{eq:CoupledMoments}) is easily solved and allows to obtain the joint distribution analytically. 
Consequently, 
\begin{multline}
p(\delta E_2,\delta \omega_2, t_2;\ \delta E_1,\delta \omega_1, t_1) = 
\mathcal{N}_{\mathbf{ \mu_1},\mathbf{ \underline{\Sigma}_1}} (\delta E_1,\delta \omega_1) \times \\ 
\mathcal{N}_{\mathbf{ \mu_2}(t_2-t_1),\mathbf{ \underline{\Sigma}_2}(t_2-t_1)} (\delta E_2,\delta \omega_2),
\end{multline}

where $\mathcal{N}_{\mathbf{ \mu},\mathbf{ \underline{\Sigma}}} (\delta E,\delta \omega)$ are bidimensionnal normal laws of mean $\mathbf{ \mu}$ and variance $\mathbf{ \underline{\Sigma}}$ and

\begin{equation}
\mathbf{ \mu_1} = 
\begin{bmatrix}
0\\
0
\end{bmatrix}
\end{equation}

\begin{equation}
\mathbf{ \mu_2} = 
\begin{bmatrix}
\delta \omega_1 \\
\delta E_1
\end{bmatrix} e^{-t}
\end{equation}

\begin{equation}
\mathbf{ \underline{\Sigma}_1} =
\begin{bmatrix}
1 & \epsilon \\
\epsilon & 1
\end{bmatrix}
\end{equation}

\begin{equation}
\mathbf{ \underline{\Sigma}_2} =
\begin{bmatrix}
1 & \epsilon \\
\epsilon & 1
\end{bmatrix} (1 - e^{-2 t}).
\end{equation}

As expected, the joint probability density function is memoryless so that it depends only on $t_2-t_1$. In the two limiting cases we recover familiar results: if $t_2 =t_1$ then $\mathcal{N}_{\mathbf{ \mu_2}(t_2-t_1),\mathbf{ \underline{\Sigma}_2}(t_2-t_1)} (\delta E_2,\delta \omega_2) = \delta (\delta E_2 - \delta E_1) \delta (\delta \omega_2 - \delta \omega_1) $; And if $t_2-t_1 \gg 1$ then  $p(\delta E_2,\delta \omega_2, t_2;\  \delta E_1,\delta \omega_1, t_1) =  p( \delta E_1,\delta \omega_1) \times p( \delta E_2,\delta \omega_2)$.

Finally, we note that for this model the statistics of the phase accumulation necessary to compute rotating frame averaging are unchanged by the correlated statistics between energy and amplitude of the driving field.

\vspace{2cm}

\bibliography{Noisy_HOM}

\end{document}